\documentclass[a4paper,10pt]{article}

\usepackage[utf8]{inputenc}
\usepackage{amsmath} 
\usepackage{amssymb}
\usepackage{graphicx}
\usepackage[margin=3cm]{geometry}
\usepackage{esint} 
\usepackage{setspace} 
\usepackage{comment}
\usepackage{epstopdf}
\usepackage{appendix}
\usepackage{url} 
\usepackage{float} 
\usepackage{booktabs} 
\usepackage{upgreek}
\usepackage{wrapfig}
\usepackage{multirow} 
\usepackage{mathtools} 
\usepackage[hidelinks]{hyperref} 
\usepackage[english]{babel}

	\usepackage{xcolor}
	\newcommand{\change}[1]{{\color{black}#1}}

	\usepackage[numbers]{natbib} 

	\DeclarePairedDelimiter\abs{\lvert}{\rvert}%
	\makeatletter
	\let\oldabs\abs
	\def\abs{\@ifstar{\oldabs}{\oldabs*}}
	\makeatother

	\newcommand{\dif}[0]{\mathrm{d}} 
	\newcommand{\pd}[2]{\frac{\partial #1}{\partial #2}}

	\newcommand{\pds}[2]{\frac{\partial^2 #1}{\partial #2 ^2}}

	\newcommand{\intn}[4]{\int_{#1}^{#2} \! #3 \, \mathrm{d} #4} 
	\newcommand{\zhat}[0]{\hat{z}}
	\newcommand{\ztilde}[0]{\widetilde{z}}

	\newcommand{\tav}[1]{\overline{#1}} 
	\newcommand{\sav}[1]{\left\langle #1 \right\rangle} 
	\newcommand{\vav}[1]{\left\langle #1 \right\rangle_{V}} 
	\newcommand{\av}[2]{\left\langle #1 \right\rangle_{#2}} 
	\newcommand{\avs}[1]{\left\langle #1 \right\rangle} 
	\newcommand{\avx}[1]{\av{#1}{x}}
	\newcommand{\avy}[1]{\av{#1}{y}}
	\newcommand{\avxy}[1]{\av{#1}{xy}}
	\newcommand{\avxz}[1]{\av{#1}{xz}}

	\newcommand{\e}[1]{\cdot 10^{#1}}
	
	\newcommand{\wu}[1]{{#1}^{+}}

	\newcommand{\urms}[0]{u_{\mathrm{rms}}} 
	\newcommand{\vrms}[0]{v_\mathrm{rms}} 
	\newcommand{\wrms}[0]{w_\mathrm{rms}} 
	\newcommand{\prms}[0]{p_\mathrm{rms}} 
	\newcommand{\dt}[0]{\Delta t} 
	
	\newcommand{\degree}[0]{^{\circ}}
	\newcommand{\plus}[0]{^{+}}
	\newcommand{\ub}[0]{U_{b}}
	\newcommand{\quan}[0]{Q}
	\newcommand{\quanref}[0]{Q^\mathrm{ref}}
	\newcommand{\kmax}[0]{k_{\max}}
	\newcommand{\tstat}[0]{t_\mathrm{stat}}
	\newcommand{\secflow}[0]{\Gamma_s}
	\newcommand{\secflowmean}[0]{\Gamma_m}
	\newcommand{\secflowturb}[0]{\Gamma_t}

	\newcommand{\upen}[0]{u_\mathrm{pen}}
	\newcommand{\vpen}[0]{v_\mathrm{pen}}	
	\newcommand{\wpen}[0]{w_\mathrm{pen}}
	\newcommand{\dretau}[0]{\Delta\retau^\mathrm{Pope}}
	\newcommand{\retauref}[0]{\retau^\mathrm{ref}}
	
	\newcommand{\uintmed}[2]{u_{#1}^{(#2)}}
	\newcommand{\rintmed}[2]{r_{#1}^{(#2)}}
	\newcommand{\pintmed}[1]{p^{(#1)}}
	\newcommand{\fdrivingintmed}[1]{f^{\mathrm{driving}(#1)}}
	\newcommand{\fIBMintmed}[2]{f_{#1}^{\mathrm{IBM}(#2)}}
	\newcommand{\pcor}[1]{\widetilde{p}^{(#1)}}
	\newcommand{\upredone}[2]{u_{#1}^{*({#2})}}
	\newcommand{\upredoneNOIBM}[2]{u_{#1}^{*({#2})\text{ no IBM}}}
	\newcommand{\upredtwo}[2]{u_{#1}^{**({#2})}}
	\newcommand{\alphas}[1]{\alpha_{(#1)}}
	\newcommand{\gammas}[1]{\gamma_{(#1)}}
	\newcommand{\zetas}[1]{\zeta_{(#1)}}

	\newcommand{\fibm}[0]{f^{\mathrm{IBM}}}
	\newcommand{\fibmx}[0]{f^{\mathrm{IBM}}_x}
	
	\newcommand{\fdriving}[0]{f^{\mathrm{driving}}}
	\newcommand{\ftot}[0]{d^{\mathrm{tot}}}

	\newcommand{\fsmooth}[0]{d_{\mathrm{smooth}}}
	
	\newcommand{\ftotsmooth}[0]{d^{\mathrm{tot}}_{\mathrm{smooth}}}
	\newcommand{\ftottextured}[0]{d^{\mathrm{tot}}_{\mathrm{textured}}}

	\newcommand{\dibm}[0]{d^{\mathrm{IBM}}}
	\newcommand{\dbulk}[0]{d^{\mathrm{bulk}}}
	\newcommand{\dtot}[0]{d^{\mathrm{tot}}}
	\newcommand{\dmeanadv}[0]{d^{\text{mean advection}}}
	\newcommand{\dturbadv}[0]{d^{\text{turbulent advection}}}
	\newcommand{\drest}[0]{d^{\mathrm{rest}}}
	\newcommand{\dsp}[0]{d_{sp}}
	\newcommand{\dcsp}[0]{dc_{sp}}

	\newcommand{\gammaf}[0]{\gamma_f}

	\newcommand{\rebulk}[0]{Re_b}

	\newcommand{\retau}[0]{Re_\tau}
	\newcommand{\retaupope}[0]{Re_\tau^\mathrm{Pope}}
	
	\newcommand{\reexp}[0]{Re_\mathrm{Bechert}}

	\newcommand{\nx}[0]{N_{x}}
	\newcommand{\ny}[0]{N_{y}}
	\newcommand{\nz}[0]{N_{z}}
	\newcommand{\dx}[0]{\Delta x}
	\newcommand{\dy}[0]{\Delta y}
	\newcommand{\dz}[0]{\Delta z}
	\newcommand{\dzw}[0]{\Delta z_{w}}
	\newcommand{\dzc}[0]{\Delta z_{c}}
	
	\newcommand{\ncg}[0]{N_{cg}}
	\newcommand{\dxplus}[0]{\wu{\dx}}
	\newcommand{\dyplus}[0]{\wu{\dy}}
	\newcommand{\dzwplus}[0]{\wu{\dzw}}
	\newcommand{\dzcplus}[0]{\wu{\dzc}}

	\newcommand{\lx}[0]{L_{x}}
	\newcommand{\ly}[0]{L_{y}}
	\newcommand{\lz}[0]{L_{z}}

	\newcommand{\sy}[0]{s_{y}}
	\newcommand{\ang}[0]{\alpha}
	\newcommand{\angbackward}[0]{\alpha_{\mathrm{backward}}}
	
	\newcommand{\hr}[0]{h_{r}}
	
	\newcommand{\ys}[0]{y_{s}}
	
	\newcommand{\splus}[0]{s\plus}
	
	\newcommand{\splusopt}[0]{s\plus_\mathrm{opt}}
	\newcommand{\nfeather}[0]{N_\mathrm{feather}}
	\newcommand{\wf}[0]{\Lambda_f} 
	\newcommand{\ngroove}[0]{N_\mathrm{groove}}
	
	\newcommand{\dr}[0]{DR}
	\newcommand{\drmax}[0]{DR_\mathrm{max}}
	\newcommand{\dc}[0]{DC}
	\newcommand{\drloss}[0]{DR_\mathrm{loss}}
	\newcommand{\tauw}[0]{\tau_w} 
	\newcommand{\utauw}[0]{u_{\tau}} 

	\newcommand{\ft}[0]{d_t}
	\newcommand{\fs}[0]{d_s}
	
	\newcommand{\lb}[0]{\left(}
	\newcommand{\rb}[0]{\right)}

\pdfoutput=1

\begin{document}



\title{Drag Reduction by Herringbone Riblet Texture in Direct Numerical Simulations of Turbulent Channel Flow}

\author{H.O.G. Benschop$^{*}$\thanks{$^*$Corresponding author. Email: H.O.G.Benschop@tudelft.nl
\vspace{6pt}}
\, and W.P. Breugem \\
\vspace{1pt} 
\em{Laboratory for Aero and Hydrodynamics, Delft University of Technology, Delft,}\\
\em{The Netherlands}\\
\vspace{6pt}
}

\maketitle

A bird-feather-inspired herringbone riblet texture was investigated for turbulent drag reduction. The texture consists of blade riblets in a converging/diverging or herringbone pattern with spanwise wavelength $\wf$. The aim is to quantify the drag change for this texture as compared to a smooth wall and to study the underlying mechanisms. To that purpose, Direct Numerical Simulations of turbulent flow in a channel with height $\lz$ were performed. The FIK-identity for drag decomposition was extended to textured walls and was used to study the drag change mechanisms.
For $\wf/\lz \gtrsim O(10)$, the herringbone texture behaves similarly to a conventional parallel-riblet texture in yaw: the suppression of turbulent advective transport results in a slight drag reduction of 2\%.
For $\wf/\lz \lesssim O(1)$, the drag increases strongly with a maximum of 73\%. This is attributed to enhanced mean and turbulent advection, which results from the strong secondary flow that forms over regions of riblet convergence/divergence. Hence, the employment of convergent/divergent riblets in the texture seems to be detrimental to turbulent drag reduction. \\



\section{Introduction} \label{sec:introduction}

Drag-reducing techniques can be of great value for fuel consumption reduction, as a significant part of the fuel used for transportation arises from drag in turbulent flows. 
Drag reduction (DR) in fluid flows can be obtained by active and passive methods.
Active methods include the use of additives \citep{white2008}, gas injection \citep{ceccio2010}, wall mass transfer \citep{kametani2011direct}, wall cooling or heating \citep{yoon2006drag, kametani2012direct, vakarelski2014leidenfrost}, wall motion \citep{quadrio2011}, wall deformation \citep{tomiyama2013direct} and electromagnetic forcing \citep{shatrov2007magnetohydrodynamic}.
Passive methods include the change of surface chemistry (e.g. superhydrophobicity \citep{rothstein2010}), elasticity (e.g. compliant walls \citep{choi1997}), shape (e.g. airfoil or ship hull shape) and texture (e.g. riblets \citep{dean2010}).
Surface roughness and surface textures have been successfully used for drag reduction by transition delay in laminar flow \citep{fransson2006delaying}, separation delay in turbulent flow over curved surfaces \citep{choi2006mechanism, son2011mechanism} and turbulence modification in turbulent flow over flat surfaces.

Of the investigated flat plate textures - such as sifted sand grains \citep{abe1990drag}, transverse square grooves \citep{wahidi2005}, dimples \citep{abdulbari2013} and V-shaped protrusions \citep{sagong2008sailfish} - the well-studied riblet texture seems most promising for turbulent DR. This texture has been found on the scales of some fast-shark skins \citep{diez2015biological} and consists of ridges or riblets aligned with the mean flow direction. With a simplified riblet geometry, a maximum DR of almost 10\% has been obtained \citep{bechert1997}.
In the search for even higher values of DR, many variations on the standard riblet geometries have been investigated \citep{dean2010}, such as hierarchical or compound riblets \citep{wilkinson1988turbulent}, riblets on a spanwise traveling surface wave \citep{li2015turbulent}, oscillating riblets \citep{wassen2008oscillating, gruneberger2013oscillating, vodop2013turbulent}, riblets in a wave-like pattern (either in phase \citep{gruneberger2012wavelike} or out of phase \citep{sasamori2014experimental}) and riblets combined with drag-reducing polymers \citep{chen2015uv}. The rationale behind these alternatives is to further reduce drag by somehow incorporating other drag-reducing methods, such as oscillating walls or polymer addition.

A little-studied alternative to the standard riblets is the herringbone riblet texture that has been found on bird flight feathers. Feathers serve several functions, including flight, thermal insulation, waterproofing and coloration (e.g. for camouflage or visual signals). The following description of the feather morphology is derived from \citet{chen2013biomimetic, chen2014flow, chen2014drag}.
A feather consists of two vanes (anterior and posterior), separated by a central supporting shaft. Each vane consists of parallel barbs attached obliquely to the shaft. The barbs are linked together by a set of finer barbs, called barbules. A microgroove is formed between neighbouring barbs. The angle between shaft and barbs is typically $\ang \approx 30 \degree$. The groove spacing $s$ remains approximately constant with $\splus \approx 20$. The groove depth decreases gradually away from the shaft. A wing is formed when several feathers are positioned next to each other with almost parallel shafts. The feather shafts are approximately parallel to the flying direction for steady forward glide.

The study of \citet{chen2014flow} is one of the first to investigate the herringbone riblet texture for turbulent drag reduction. Inspired by bird flight feathers, they designed a riblet texture with two typical features that differ from the shark-skin riblets. First, the riblets were arranged in a converging/diverging or herringbone pattern. Second, the riblet height or groove depth changed gradually. Using laser engraving and replica molding, the researchers manufactured such bio-inspired herringbone-riblet skins. They covered the inner wall of a test pipe with these skins and measured a DR of up to 20\%, twice that of optimal standard riblets. Although this seems promising, it has not been reproduced yet, and it is unclear for which texture and flow parameters drag is maximally reduced.

\citet{sagong2008sailfish} investigated a rather comparable geometry, namely the V-shaped protrusions that were found on the sailfish skin. In a comprehensive experimental and numerical study, they found a few cases for which drag was decreased slightly ($\sim 1$\%), although within the experimental uncertainty. 
\change{
The herringbone riblet texture also resembles the vortex generators that have been used to reduce drag by flow-separation delay \citep{lin2002review} or transition delay \citep{shahinfar2012revival}. Furthermore, several studies confirm that roughness on a bird wing contributes to separation control \citep{bushnell1991nature, lilley1998study, bokhorst2015feather}.}

Herringbone riblet textures have been studied for several other reasons, such as for heat-transfer enhancement \citep{gao2001heat, fang2015highly} and mixing of laminar flows in microchannels \citep{stroock2002chaotic}.
The hot-wire study by \citet{koeltzsch2002} was initiated by the observation of convergent and divergent riblet patterns on the shark skin near sensory organs, possibly used for local flow-noise reduction.
\citet{nugroho2013large} revisited the same texture and conducted a parametric study using the hot-wire technique. They were motivated by the potential use for passive flow control and disruption of large-scale coherent motions. They proposed that the herringbone riblets induce large-scale counter-rotating vortices, giving rise to large-scale spanwise periodicity in the boundary layer. 
The suggested streamwise counter-rotating vortices have recently been visualized in a flat-plate laminar boundary layer over convergent riblets \citep{nadesan2014quasi}.
So, the motivation for herringbone riblet studies has been drag reduction, heat-transfer enhancement, mixing improvement, flow control and large-scale flow structures manipulation.


The study by \citet{nugroho2013large} is a manifestation of the recent interest in surfaces that manipulate the whole boundary layer. There is much evidence that DR techniques that rely on near-wall flow manipulation are less effective at higher Reynolds numbers: $DR$ degrades with increasing $Re$ \citep{iwamoto2002reynolds, iwamoto2005friction, spalart2011drag, gatti2013performance}. That has partially motivated the study of rough or textured surfaces that also impact the outer part of the turbulent boundary layer by means of large-scale secondary flows. These secondary flows have been observed over 
spanwise-varying longitudinal bedforms \citep{wang2006time}, 
a regular array of cubes \citep{reynolds2007spanwise},
a bed with two parallel lanes of different roughnesses \citep{vermaas2011lateral},
the irregular surface of a replica of a damaged turbine blade \citep{barros2014observations}, 
streamwise strips of elevated roughness \citep{vanderwel2015effects} and 
the aforementioned herringbone riblet texture \citep{koeltzsch2002, nugroho2013large}.
Interestingly, \citet{schoppa1998large} obtained 20\% DR in Direct Numerical Simulations of turbulent channel flow with imposed large-scale counter-rotating streamwise vortices.

The aim of this paper is to quantify how and why drag is changed by a herringbone texture as compared to a smooth wall, using Direct Numerical Simulations (DNSs). 
The numerical methods are described in section \ref{sec:numerical_methods}.
Section \ref{sec:DR_quantification} explains how drag reduction is quantified.
Section \ref{sec:validation} validates the simulated drag of smooth walls, parallel riblets aligned with the mean flow, and parallel riblets in yaw.
Section \ref{sec:herringbone_riblets} examines the herringbone texture with use of a parametric study.
A drag decomposition is derived and applied in section \ref{sec:fik}.
A discussion of the results is presented in section \ref{sec:discussion}, followed by the main conclusions and an outlook in section \ref{sec:conclusions}.

\section{Numerical Methods} \label{sec:numerical_methods}

In this study, Direct Numerical Simulations (DNSs) of incompressible turbulent flow were performed. An overview of all simulations with the corresponding parameters can be found in \autoref{sec:simulation_parameters}. This section outlines the method that was used, describing successively the notation conventions, flow domain, texture, grid, numerical code and averaging.

\textbf{Notation}
In what follows, dimensional variables are denoted by an asterisk $^{*}$. Variables without that asterisk are nondimensionalized using the domain height $\lz^*$ and the bulk velocity $\ub^*$, such that $\lz = 1$ and $\ub = 1$. Note that $\ub^*$ is a constant, as simulations were performed at fixed mass flow rate. \change{The constant bulk Reynolds number is defined as $\rebulk = \ub^* \lz^* / \nu^*$, with kinematic viscosity $\nu^*$.}
The superscript $\plus$ is used for nondimensionalization with $\nu$ and $\utauw = \sqrt{\tauw/\rho}$, with wall shear stress $\tauw$ and fluid density $\rho$. Nondimensionalization for textured walls uses the viscous wall units derived from the smooth-wall simulation with the same $\rebulk$.

\textbf{Domain}
The flow domain is a plane channel, bounded by two horizontal walls. At the channel walls, no-penetration and no-slip boundary conditions are applied, whereas periodic boundary conditions are used in the streamwise and spanwise directions. The streamwise, spanwise and wall-normal coordinates are denoted by $x$, $y$ and $z$ with the corresponding velocity components $u$, $v$ and $w$.

The domain is specified by its length $\lx$, width $\ly$ and height $\lz$. For a good comparison, the domain size should ideally be the same for all simulations. However, slight size variation was needed to fit an integer number of texture periods in the streamwise and spanwise directions, or to ensure that the number of grid cells complies with the parallel-computing algorithm. In general, all domains are approximately of size $(4.0 \times 2.5 \times 1)$, which is considered to be large enough to obtain reliable statistics for several reasons.
First, it is comparable to the domain size used by other researchers \citep{moser1999direct, breugem2005, orlandi2006turbulent, vreman2014statistics}. In addition, it is full-span, as opposed to the recently reconsidered minimal-span channels \citep{chung2015fast}. Finally, it is larger than the moderate box of size $(\pi \times \pi/2 \times 1)$ that is large enough to reproduce the one-point statistics of larger boxes \citep{lozano2014effect}.






\textbf{Texture}
Textures are applied to the inside of both channel walls to enforce symmetry in the mean flow \citep{garcia2011hydrodynamic}. Unless stated otherwise, the top wall texture is the bottom wall texture mirrored in the centerline plane.
To simulate flow over a non-smooth surface, two methods can be adopted: coordinate transformation or the Immersed Boundary Method (IBM) \citep{orlandi2006turbulent}. Both methods have been applied to simulate turbulent flow over riblet walls \citep{choi1993DNS, goldstein1995direct}.
We used an IBM similar to the one employed by \citet{breugem2005} and \citet{pourquie2009some}, which is based on \citet{fadlun2000combined}. The IBM forcing is direct, i.e. a forcing term is added to the discretized equations. Appendix \ref{sec:IBM} provides case-specific details.


\textbf{Grid}
%
The chosen IBM allows the use of a simple staggered Cartesian grid. The number of grid cells in the three Cartesian directions is denoted by $\nx$, $\ny$ and $\nz$. The grid is uniform in the horizontal directions, so the grid spacings $\dx$ and $\dy$ are constant. 
In the wall-normal direction, three zones are distinguished, namely the roughness regions near the two walls and the remaining part of the channel. 
In the roughness region, which extends from the wall to one grid cell above the maximum texture height, the vertical grid spacing $\dzw$ is constant.
In the remaining part of the channel, grid stretching is applied using a cosine function that is symmetrical with respect to the channel centerline. The maximum vertical grid spacing occurs at the centerline and is called $\dzc$.


The grid-cell size is important to correctly resolve small-scale fluid motions. The recommendations of \citet{vreman2014comparison} for finite difference codes were followed, namely $\dxplus = 4.4$, $\dyplus = 2.9$, $\dzwplus = 0.49$ and $\dzcplus = 2.2$ as maximum grid spacings for smooth-wall turbulent flows.
For textured walls, the spanwise grid spacing was reduced to $\dyplus \approx 1.0$.
For all investigated textures, one simulation at a double spanwise and/or streamwise resolution was performed. Particular attention was paid to the resulting drag change, which was marginal in all cases. As only one simulation at a higher resolution was performed for each texture, grid independence of the results cannot be claimed. However, as the grid resolution is relatively high and about the same in all cases, comparison of results is still justified.

\textbf{Code} 
%
The incompressible Navier-Stokes equations and continuity equation were solved at fixed bulk velocity:
\begin{align}
\pd{u_i}{t} + \pd{u_i u_j}{x_j} & = 
-\pd{p}{x_i} + \frac{1}{\rebulk} \pds{u_i}{x_j} + \gammaf \fdriving \delta_{i1} + \fibm_i, \label{eq:Navier_Stokes_equations} \\
\pd{u_j}{x_j} & = 0,
\end{align}
where the Einstein summation convention for repeated indices is used. Here $u_i$ represents one component of the velocity vector, $t$ time, $x_j$ a spatial coordinate, $p$ the pressure and $\fdriving$ the spatially uniform forcing term to obtain a constant bulk velocity. The phase-indicator function $\gammaf$ is defined at grid points of the streamwise velocity. It equals 1 in fluid and 0 in solid obstacle volume to ensure that only fluid experiences the bulk forcing that drives the flow. The Kronecker-delta function $\delta_{i1}$ guarantees that fluid is driven in the streamwise direction with $i = 1$. \change{\autoref{sec:numerical_code} describes how $\fdriving$ is calculated. The IBM forcing $\fibm_i$ is a body force that models the (drag) force that the texture exerts on the flow (see \autoref{sec:IBM} for more details)}.
 

These equations are discretized using the finite-volume method combined with a pressure-correction scheme \citep{ferziger2002computational}. Fluxes or stresses at the cell faces are evaluated using linear interpolation, i.e. a central-differencing scheme is used. Time-integration is performed using a fractional-step method that consists of three steps. For the pressure, which is staggered in time with respect to the velocities, a Crank-Nicolson scheme is used. All other terms are advanced in time using a three-step Runge-Kutta method \citep{wesseling2001principles}. This discretization procedure yields a Poisson equation, which is solved using a non-iterative FFT-based solver. Specifically, FFTs are applied to the horizontal directions and the resulting tridiagonal system is solved using Gaussian elimination. \change{More details about the time advancement at fixed bulk velocity can be found in \autoref{sec:numerical_code}.}

\textbf{Averaging}
For computation of flow statistics, a unit-cell average was stored each 100 timesteps. Like in crystallography, a unit cell is the smallest unit of volume that builds up the entire texture by translation. It extends vertically from bottom to top wall. For smooth walls, its size in grid cells is $1 \times 1 \times \nz$. In a unit-cell average, the data of all unit cells are reduced to an average in one unit cell.

In addition to this unit-cell average during the computations, temporal and spatial averages were performed afterwards. Let $\phi = \phi(x,y,z,t)$ represent an arbitrary flow variable. The following averages were used:
\begin{align}
\tav{\phi}				& = \frac{1}{T} \int_{\tstat}^{\tstat + T} \phi \: \dif t, \\
\av{\phi}{x_{i}}		& = \frac{1}{L_{x_i}} \int_{0}^{L_{x_i}} \phi \: \dif x_i, \\
\vav{\phi}		  		& = \frac{1}{V} \int_{V}\! \phi \: \dif V.
\end{align}
Here $\tav{\phi}$ is a time average over the statistically stationary part of the signal $\phi$ (which starts at $\tstat$ and has duration $T$).
An average over one spatial coordinate is denoted by $\av{\phi}{x_{i}}$. For instance, $\av{\phi}{x}$ is a streamwise average. A similar notation is adopted for an average over two spatial coordinates. For example, $\avxy{\phi}$ is a streamwise and spanwise average.
Finally, $\vav{\phi}$ represents an average over the entire volume $V = \lx \ly \lz$. For instance, the bulk velocity is defined by $\ub = \vav{u}$.



\section{Drag Reduction Quantification} \label{sec:DR_quantification}


This section describes the quantification of drag reduction. It explains under what conditions the smooth- and textured-wall flows are compared, with special attention for flow generation, the definition of a reference case and the formulation of the benefit of drag reduction.


In general, channel flow is generated by either a constant flow rate (CFR), a constant pressure gradient (CPG) or a constant power input (CPI) \citep{frohnapfel2012money}. The obtained drag reduction depends slightly on the choice of CFR, CPG or CPI. In this study, the flow was generated by CFR.


The quantification of drag reduction requires the definition of a reference case compared to which drag is reduced or increased. Special attention should be paid to the Reynolds number and channel height, as drag depends heavily on both of them. 
Conceptually, this study compares two channels with the same fluid, the same flow rate per unit of spanwise width and the same outer dimension $\lz^*$. 
With $\ub^* \lz^*$ being the volumetric flow rate per unit width, the first two conditions imply that $\rebulk$ is the same for both flows.
The third condition guarantees that both channels are geometrically identical, apart from the texture that is applied to the inside of the channel walls in one case. It implies that the wall location is not adjusted to compensate for the texture volume. The fluid volume for the case with textured walls is thus slightly less than that of the smooth-wall case. This is a conservative choice: drag reduction cannot result from an increased fluid volume or a locally increased channel height \citep{daschiel2012flow, mohammadi2013groove}.

This paper uses the drag change $\dc$ as a quantifying parameter. To account for possible differences in domain width and length, $\dc$ was computed from the time-averaged drag force per unit volume. As the driving term balances the total drag, the instantaneous drag force per unit volume 
$f_d^* = \rho^* \lb \ub^* \rb^2 \lb \lz^* \rb^{-1} V^{-1} \int_V \gammaf \fdriving \dif V$.
Since the dimensional prefactor is equal for the smooth- and textured-wall channel flows, the drag change is given by
\begin{equation}
\begin{aligned}
\dc & = 
\frac{ \;\;\; \left. \tav{ \vav{\gammaf \fdriving} } \, \right|_\mathrm{textured} \;\;\; }
     { \left. \tav{ \vav{\gammaf \fdriving} } \, \right|_\mathrm{smooth} } - 1 \\
& \equiv \frac{ \;\;\; \ftottextured \;\;\; }{ \ftotsmooth } - 1.
\end{aligned}
\label{eq:DC}
\end{equation}
%
%
%
%
It measures the increase of the driving force that is required to maintain a given flow rate. The drag reduction $DR = -\dc$, so drag is reduced in case $\dc$ is negative.


The thus computed drag change is supplemented by a 95\% confidence interval. The error in $\dc$ can be attributed to the uncertainty in the drag computed for both the textured and smooth wall. For ease of notation, define $\ft \equiv \ftottextured$ and $\fs \equiv \ftotsmooth$. Let $u_\phi$ for now denote the uncertainty in $\phi$. 
%
%
Given the independence of $\ft$ and $\fs$, the uncertainty in $\dc$ follows from the law of error propagation:
\begin{equation}
\begin{aligned}
u_{\dc}^2 =
\lb \frac{u_{\ft}}{\fs} \rb^2 + 
\lb \frac{\ft u_{\fs}}{\fs^2} \rb^2.
\end{aligned}
\end{equation}
%
The uncertainties in $\ft$ and $\fs$ were computed using the method outlined by \citet{hoyas2008reynolds}. It accounts for correlation in the drag time signal. The thus obtained error bar only results from the finite simulation time. Errors of other origins (e.g. discretization errors) were not considered.

\begin{table*}[t!]
  \centering
  \begin{minipage}[t]{1\textwidth}
  \caption{Validation of drag and flow statistics of smooth-wall simulations. 
  The relative deviation of $\retau$ from that predicted by Pope's relation is given by  		$\dretau$ (\autoref{eq:dretau}).
  Statistics of $U$, $\urms$, $\vrms$, $\wrms$ and $\prms$ are compared with simulations by \citet{vreman2014comparison, vreman2014statistics} at $\retauref = 180$ and $590$.
  The relative difference between current and reference flow statistic $\quan$ is measured with the root-mean-square relative deviation $\delta \quan$ (\autoref{eq:deltaquan}).}
  \resizebox{\columnwidth}{!}{
    \begin{tabular}{ l l l l l l l l l }
	\hline
$\rebulk$	 & $\retau$	 & $\dretau \, [\%]$	 & $\retauref$	& $\delta U \, [\%]$	 & $\delta \urms \, [\%]$	 & $\delta \vrms \, [\%]$	 & $\delta \wrms \, [\%]$	 & $\delta \prms \, [\%]$ \\ \hline
5500	 & 175.0	 & -0.6	 & 180	 & 0.58	 & 1.09	 & 1.39	 & 1.52	 & 3.98 \\ 
11000	 & 320.9	 & -1.0	 & - 	 & -	 & -	 & -	 & -	 & - \\ 
22000	 & 587.4	 & -1.5	 & 590	 & 0.55	 & 1.85	 & 0.82	 & 0.88	 & 0.94 \\ 
    \hline
    \end{tabular}%
    }
    \label{tab:smooth_validation}
    \end{minipage}
\end{table*}%

\section{Validation} \label{sec:validation}

The numerical methods were validated with simulations of smooth walls, parallel riblets, and parallel riblets in yaw, as described in the following subsections.


\subsection{Smooth Wall} \label{sec:smooth_wall}

Smooth-wall DNSs were performed at three bulk Reynolds numbers, namely 5500, 11000 and 22000. \change{\autoref{tab:smooth_validation}} shows the corresponding friction Reynolds number $\retau = \utauw \delta/ \nu$, where $\delta = \lz/2$ is the half-channel height. Compared to recent DNSs reaching $\retau = 4000$ \citep{bernardini2014velocity}, $\retau \approx 4200$ \citep{lozano2014effect} and $\retau \approx 5200$ \citep{lee2015direct}, the simulations in this study are considered to be standard. Therefore, this subsection suffices to validate drag and flow statistics.

Drag is validated by comparing $\retau$ with the value predicted by an approximate relation $\retaupope = 0.09 \rebulk^{0.88}$ \citep{pope2000turbulent}. The deviation of $\retau$ from $\retaupope$ is quantified using
\begin{equation}
\begin{aligned}
\dretau = \frac{\retau}{\retaupope} - 1.
\end{aligned}
\label{eq:dretau}
\end{equation}
\change{\autoref{tab:smooth_validation}} lists $\dretau$ for all smooth-wall simulations. Although the relation for $\retaupope$ is approximate, good agreement is obtained for all $\rebulk$.

Flow statistics of $U$, $\urms$, $\vrms$, $\wrms$ and $\prms$ were compared with simulations at $\retau = 180$ \citep{vreman2014comparison} and $\retau = 590$ \citep{vreman2014statistics}. Here $U = \avxy{\tav{u}}$, $\urms = ( \langle \, \tav{u^2} \, \rangle_{xy} - \avxy{\tav{u}}^2 )^{1/2}$ and similarly for the other root-mean-square quantities. Let $\quan(z)$ be one of these statistics, then the root-mean-square relative deviation
\begin{equation}
\begin{aligned}
\delta\quan = \sqrt{\av{\lb \frac{\quan(z) - \quanref(z)}{\quanref(z)} \rb^2}{z}}
\end{aligned}
\label{eq:deltaquan}
\end{equation}
is used to quantify the difference between $\quan$ (current) and $\quanref$ (reference).
It was computed after piecewise cubic spline interpolation of $\quan$ and $\quanref$ to a uniform grid, $z\plus(k) = k$ for integers $1 \le k \le \kmax$, with $\kmax = 175$ at the lowest and $\kmax = 587$ at the highest $\rebulk$. 
\change{\autoref{tab:smooth_validation}} shows $\delta\quan$ for the five flow quantities. The root-mean-square relative deviation is smaller than 1\% for the mean velocity and smaller than 2\% for the root-mean-square fluctuations. The somewhat larger value for $\prms$ at the lowest $\rebulk$ is attributed to the difference between $\retau$ and $\retauref$.

\begin{figure*}[t!]
\begin{minipage}[t]{1\linewidth}
\centering
\includegraphics[width=\textwidth]{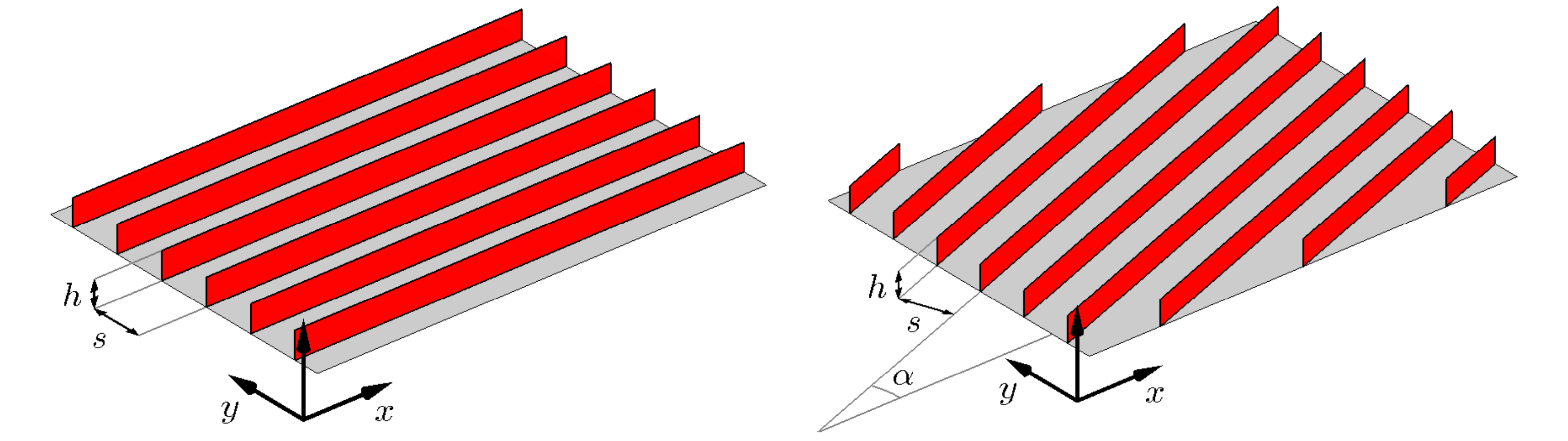}
\end{minipage}
\caption{Parallel blade riblet texture.
\textbf{Left:} Bird's-eye view of the texture aligned with the mean flow, showing six unit cells in the spanwise direction.
\textbf{Right:} Bird's-eye view of the texture in yaw with yaw angle $\alpha$, showing six unit cells in the spanwise and three in the streamwise direction.} 
\label{fig:parallel_riblet_geometry}
\end{figure*}

\subsection{Parallel Riblets} \label{sec:parallel_riblets}

A parallel riblet texture consists of riblets aligned with the mean flow direction and can reduce turbulent drag up to almost 10\% \citep{bechert1997}. An overview of previous research can be found elsewhere \citep{dean2010}. A thorough DNS study has been performed quite recently \citep{garcia2011drag, garcia2011hydrodynamic, garcia2012scaling}.

This paper investigates the blade riblet texture (see \autoref{fig:parallel_riblet_geometry}). The blades have zero thickness, spacing $s$ and height $h$ with $h/s = 0.5$. In a small parametric study, mainly $\splus$ and $\rebulk$ were varied (see also \autoref{tab:simulation_parameters}). The grid resolution (specified in wall units) is about the same for all cases.

\begin{figure*}[t!]
\begin{minipage}[t]{\linewidth}
\centering
\includegraphics[width=\textwidth]{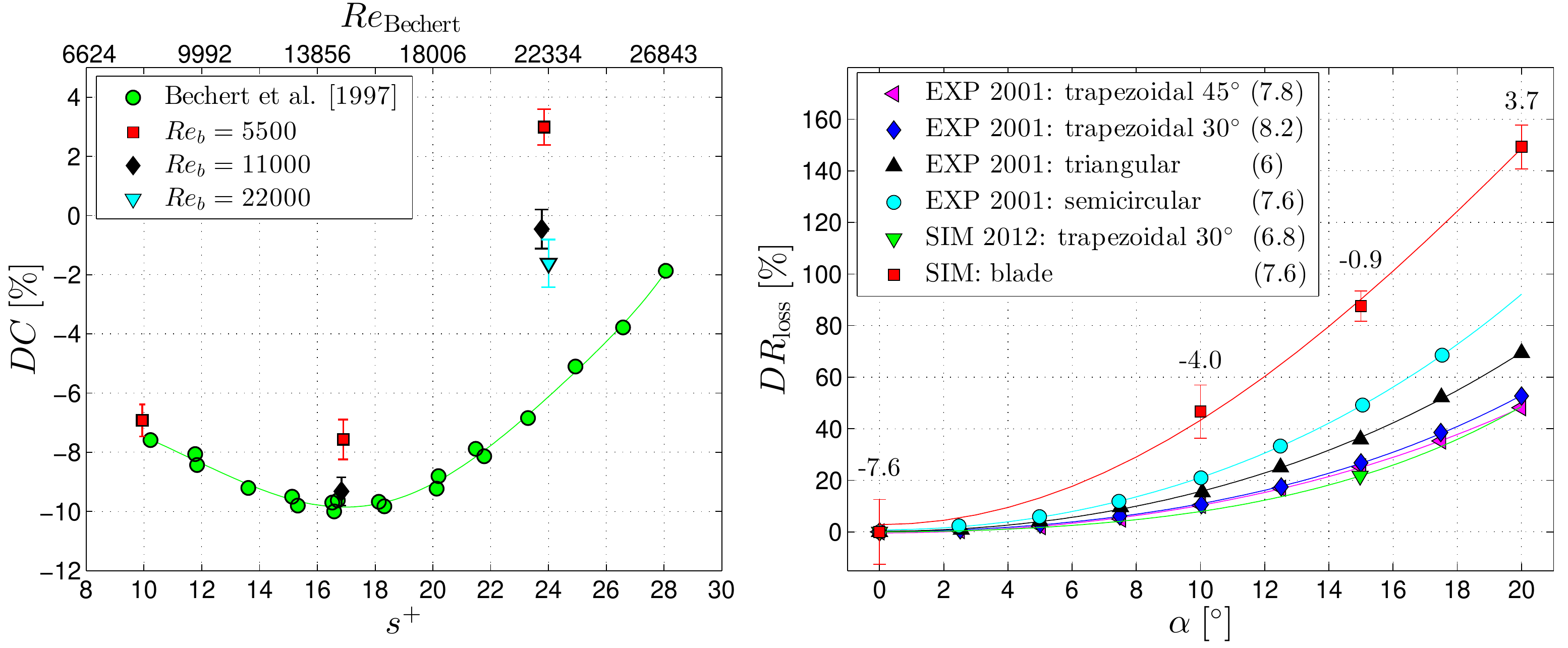}
\end{minipage}
\caption{Validation of drag change for the parallel blade riblet texture.
\textbf{Left:} Drag change as function of riblet spacing in wall units (no yaw). Numerical results at three different Reynolds numbers are compared with experimental data of \citet{bechert1997}. The experimental bulk Reynolds number is denoted by $\reexp$.
\textbf{Right:} Drag reduction loss (\autoref{eq:DR_loss}) as function of yaw angle for different riblet geometries at $\splus = 17$. \change{EXP 2001 are experiments by \citet{hage2001yaw} at $\rebulk \approx 14900$. SIM 2012 are DNSs by \citet{gruneberger2012wavelike} at $\rebulk = 5750$.} SIM are the current simulations. The numbers in parentheses in the legend represent the maximum drag reduction $\drmax$ in percent. The numbers in the figure represent the drag change $\dc$ in percent for the simulated blade riblet geometry.
\newline}
\label{fig:drag_parallel_riblets}
\end{figure*}

\autoref{fig:drag_parallel_riblets} (left) compares the simulated drag change as function of $\splus$ with experiments performed by \citet{bechert1997}. The top axis shows the experimental bulk Reynolds number $\reexp$, which is based on the horizontal channel width and the average velocity between the test plates. In the experiments only the Reynolds number was varied, whereas in the present numerical study both the Reynolds number and the riblet spacing were varied.


\change{The drag change varies slightly with Reynolds number for fixed $\splus$: the drag at $\rebulk = 5500$ is higher than at $\rebulk = 11000$ and $22000$.} The approximate overlap of the data points at the two highest Reynolds numbers (for $\splus = 24$) suggests a low-Reynolds-number effect, which is underpinned by the observation that drag reduction data below $\reexp \approx 10000$ deviated more and more from previous high-Reynolds-number data \citep{bechert1997}. 
The deviation of $\dc$ at $\rebulk = 5500$ from that at higher Reynolds numbers is also larger at larger $\splus$, which might be explained by riblet height increase. For $\rebulk = 5500$, an increase of $\splus$ from 10 to 24 is accompanied by a decrease of $\delta/h$ from 35 to 15. The blades protrude farther into the channel, which is presumably detrimental to drag reduction. This is supported by the finding that, for $\delta/h \lesssim 50$, the effect of roughness extends across the boundary layer and the original wall flow dynamics is changed significantly \citep{jimenez2004rough}.

The simulations capture the experimental trend quite well when $\rebulk$ is close to $\reexp$. Especially good agreement is obtained around and below the optimum spacing. At $\splus = 17$, a maximum drag reduction of 9.3\% is achieved, which is very close to the 9.9\% of the experiments. The difference is slightly larger at $\splus = 24$: there the total drag is reproduced within 5\%. A higher reproduction accuracy probably requires a combined experimental/numerical study, a more extensive grid resolution study, the incorporation of blade thickness and an analytical correction of momentum fluxes near the riblet tips to resolve the high gradients there. However, the accuracy demonstrated here is sufficient for the herringbone riblet simulations, as drag differences for these were found to be much larger than 5\%.




\subsection{Parallel Riblets in Yaw} \label{sec:parallel_riblets_yaw}

Among other factors, the performance of riblets deteriorates in yaw, i.e. when they are not aligned with the mean flow direction. An overview of past research is given by \citet{koeltzsch2002}. The study of \citet{hage2001yaw} shows the significant influence of riblet geometry and spacing $\splus$. Simulations of turbulent flow over parallel riblet textures in yaw are rare in the literature. \change{\citet{gruneberger2012wavelike} have performed DNSs at $\rebulk = 5750$ ($\retau = 180$) for trapezoidal grooves.} The driving pressure gradient was rotated such that the streamwise direction no longer coincides with the $x$-direction. Good agreement with experimental data was obtained.

This paper investigates the blade riblet texture in yaw (see \autoref{fig:parallel_riblet_geometry}) with $h/s = 0.5$ and $\splus = 17$. The yaw angle $\ang$ was varied for fixed $\rebulk = 5500$. The horizontal grid resolution was comparable to that used for aligned parallel riblets: $\dxplus \le 4.1$ and $\dyplus \le 1.1$ for all cases. As the trick of driving-pressure-gradient rotation cannot be applied for herringbone-riblets simulations, the parallel riblet texture was rotated with respect to the grid. That required a different Immersed Boundary Method (IBM), as the blades are not anymore aligned with the Cartesian directions (see Appendix \ref{sec:IBM} for details).


\autoref{fig:drag_parallel_riblets} (right) shows $\dc$ as function of $\ang$ (indicated by the numbers in the figure). Clearly, the drag reduction diminishes with increasing yaw angle, as is expected from the literature.
Unfortunately, to the best of the authors' knowledge, no experimental data of blade riblets in yaw is available. For comparison of the simulated results with experimental data for other riblet geometries, the drag reduction loss $\drloss$ is introduced \citep{hage2001yaw}:
\begin{equation}
\begin{aligned}
\drloss(\ang, \splus) = 
\frac{ \drmax - \dr(\ang, \splus) }{ \drmax },
\end{aligned}
\label{eq:DR_loss}
\end{equation}
%
%
where $\dr = -\dc$, and $\drmax$ is the maximum drag reduction that can be obtained with a given geometry: $\drmax\!=\dr(\ang\!=\!0, \splus\!=\!\splusopt)$. It was assumed that $\drmax\!=\dr(\ang\!=\!0, \splus\!=\!17)$ for the simulations.

\autoref{fig:drag_parallel_riblets} shows the drag reduction losses as function of yaw angle for different riblet geometries at $\splus = 17$. \change{The experimental data of \citet{hage2001yaw} (at $\rebulk \approx 14900$) and the numerical data of \citet{gruneberger2012wavelike} (at $\rebulk = 5750$) are included.} The values of $h/s$ are geometry dependent, namely $h/s = 0.5$ for the trapezoidal and blade, $h/s = 0.7$ for the semicircular, and $h/s = 1$ for the triangular geometries.


\change{
The dependence of $\drloss$ on $\ang$ is different for each geometry, which might be explained by differences in $h/s$ and riblet shape \citep{hage2001yaw}. Of the two trapezoidal geometries, the one with the sharpest tip ($30\degree$) is slightly more sensitive to misalignment. The triangular geometry has the largest tip angle of about $54\degree$, but its drag reduction loss is nevertheless larger than for the trapezoidal geometries, presumably because of its larger height ($h/s = 1$). Of the experimental data, the semicircular geometry is most sensitive to yaw, likely because of its larger height ($h/s = 0.7$) and sharp tip.
}

The simulation results suggest that the blade riblet geometry is more susceptible to yaw than any of the other geometries. This is likely not a low-Reynolds-number effect, as the simulations at \change{$\rebulk = 5750$} by \citet{gruneberger2012wavelike} reproduced well the experimental data (see the figure). Instead, it might be ascribed to the sharper riblet tips (and the associated pressure drag increase) and the broader riblet valleys (and the associated increase of sloshing) \citep{hage2001yaw}.

\begin{figure*}[t!]
\begin{minipage}[t]{\linewidth}
\centering
\includegraphics[width=\textwidth]{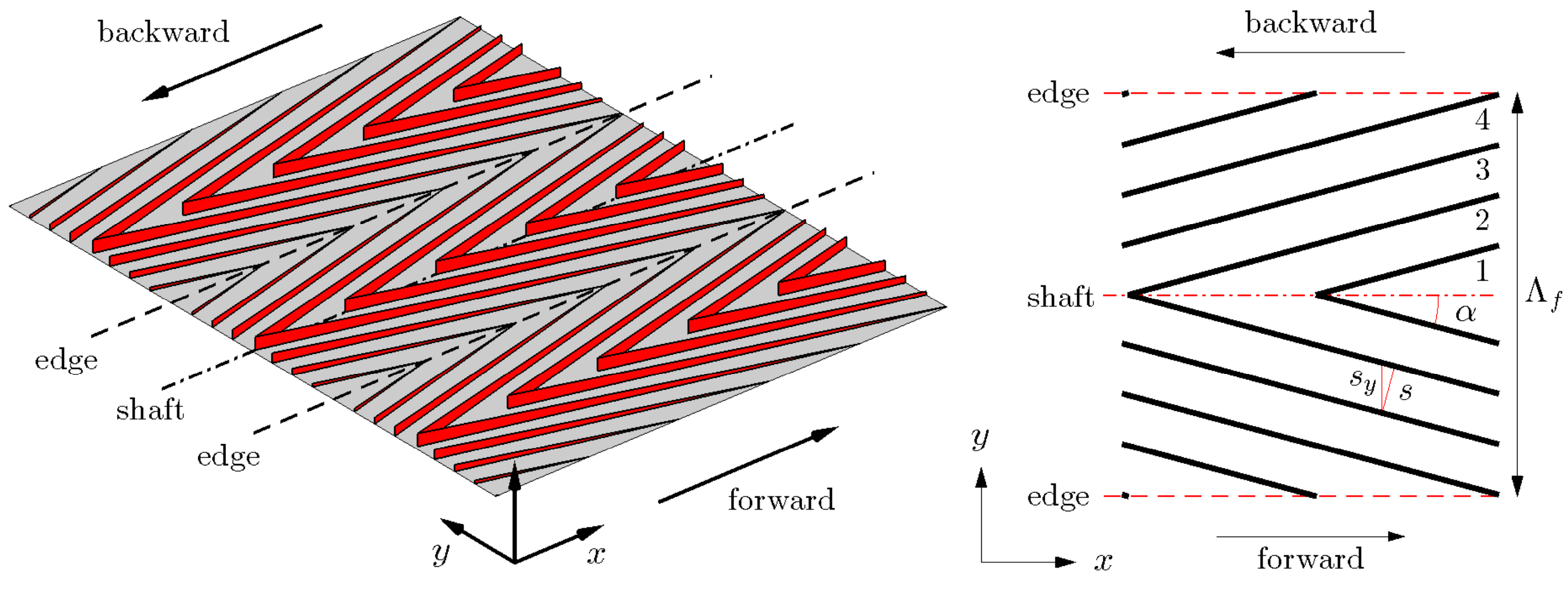}
\end{minipage}
\caption{Herringbone blade riblet texture with 4 grooves per feather half ($\ngroove = 4$), and angle between shaft and riblets of 15 degrees ($\ang = 15\degree$).
\textbf{Left:} Bird's-eye view of the texture, showing five unit cells in the streamwise and three unit cells (or feathers) in the spanwise direction. The edges and shaft of the middle feather are shown. Backward flow over this texture is equivalent to forward flow over a texture with $\ang = 165\degree$.
\textbf{Right:} Top view of the texture, showing two unit cells in the streamwise and one unit cell in the spanwise direction. The feather width is $\wf$. Four grooves on one feather half are numbered.}
\label{fig:model_herringbone_texture}
\end{figure*}

\section{Herringbone Riblets} \label{sec:herringbone_riblets}





\subsection{Texture Description}

The feather texture (described in \autoref{sec:introduction}) is modeled as shown in \autoref{fig:model_herringbone_texture}, fairly similar to the spatial 3D (s-3D) texture proposed in \citet{chen2014flow}. It consists of $\nfeather$ feathers placed in parallel such that their edges touch. 
The modeled feather has no physical shaft, although the term `shaft' is employed to denote the symmetry axis of a feather.
The barbs are modeled as blade riblets with zero thickness. As opposed to the s-3D texture that consists of sawtooth riblets, the present study uses blades because of their superior drag-reducing performance in the conventional riblet texture.
The angle between the positive $x$-direction and the blades is called $\ang$, which is restricted to $0 \leq \ang < 180 \degree$. The shortest distance between the blades is  $s$. 
Between two neighboring blades a groove forms, which has spanwise blade spacing $\sy = s/\cos(\ang)$. $\ngroove$ is the integer number of spanwise blade spacings $\sy$ that fits in one feather half-width (see also \autoref{fig:model_herringbone_texture}).
The feather width or spanwise texture wavelength $\wf = 2 \ngroove \sy$.
The riblet height $\hr$ decreases linearly with distance to the shaft. Let $h$ represent the riblet height at the shaft and $\ys$ the $y$-coordinate of the shaft, then $\hr$ is given by 
\begin{equation}
\frac{\hr(y)}{h} = 
1 - \abs{ \frac{y - \ys}{\wf/2} }, 
\quad \text{for } -\frac{1}{2} \leq \frac{y - \ys}{\wf} \leq \frac{1}{2}.
\label{eq:herringbone_riblet_height}
\end{equation}
Note that \citet{koeltzsch2002} and \citet{nugroho2013large} used herringbone riblets of constant height.
%
%
The texture was implemented using the same IBM that was used for blade riblets in yaw; only the texture indicator functions were different.


As \autoref{fig:model_herringbone_texture} shows, a difference is made between forward and backward flow. `Forward' is used for bulk flow in the positive, `backward' indicates flow in the negative $x$-direction. Forward flow over a texture with angle $\ang$ is the same as backward flow over a texture with angle $\angbackward = 180\degree - \ang$. Using this trick, textures with angle $\ang$ and $\angbackward$ can be compared to study the influence of mean flow direction on drag.

\begin{figure*}[t!]
\begin{minipage}[t]{\linewidth}
\centering
\includegraphics[width=\textwidth]{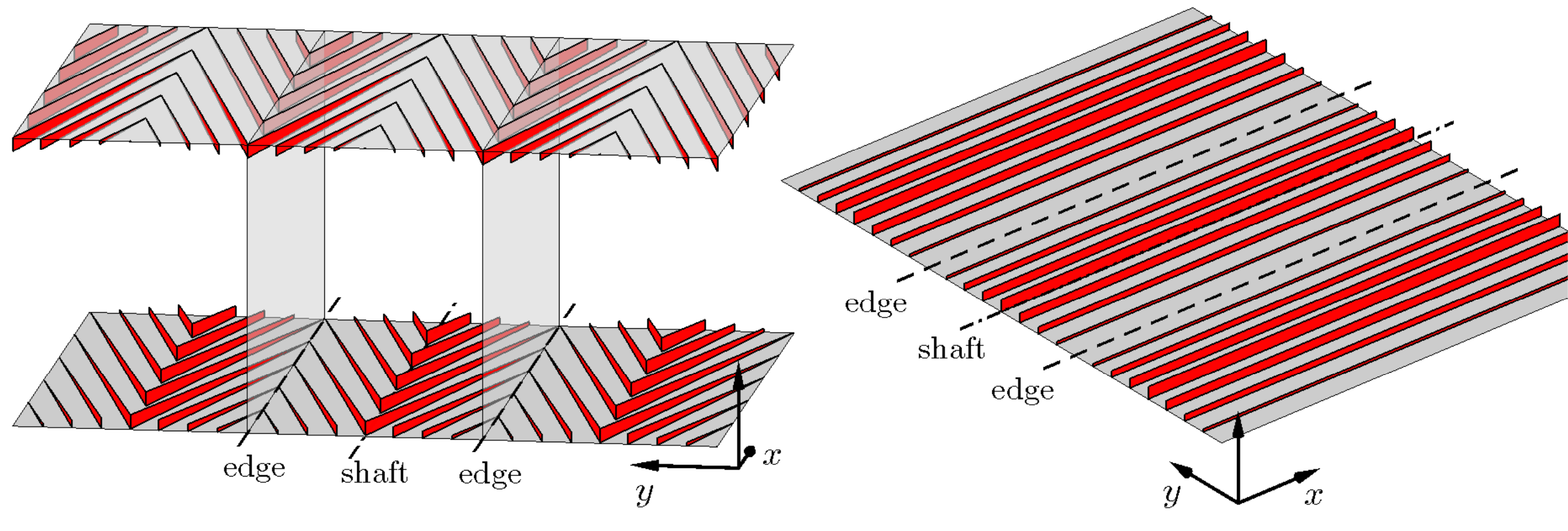}
\end{minipage}
\caption{Herringbone riblet texture variations.
\textbf{Left:} Bird's-eye view of a shifted texture variation with $\ngroove = 4$ and $\ang = 15\degree$. `Shifted' refers to the spanwise shift of the top wall texture by half a spanwise texture wavelength as compared to the bottom wall. The two vertical planes demarcate the middle unit cell. Note that the distance between bottom and top wall is not to scale.
\textbf{Right:} Bird's-eye view of a parallel texture variation with $\ngroove = 4$ and $\ang = 0\degree$, showing three unit cells (or feathers) in the spanwise direction.}
\label{fig:model_herringbone_texture_variations}
\end{figure*}

In addition to the standard herringbone texture, two texture variations were considered. 
The first variation is called `shifted'. It differs from the standard herringbone texture by a spanwise shift of the top wall texture by half a spanwise texture wavelength as compared to the bottom wall (see \autoref{fig:model_herringbone_texture_variations}). This shifted texture was investigated for its ability to generate the drag-reducing secondary flow that is described by \citet{schoppa1998large}, namely one that extends from the bottom to the top wall.
%
%
The second variation is a riblet texture with $\ang = 0\degree$ (see \autoref{fig:model_herringbone_texture_variations}), which results in a parallel blade riblet geometry with blade height variation in the spanwise direction given by \autoref{eq:herringbone_riblet_height}. This texture does not suffer from yaw and the resulting pressure drag, but it still might give rise to secondary flows.

\change{

The herringbone texture and its variations have been subjected to a parametric study (see also \autoref{tab:simulation_parameters}). According to \citet{chen2014flow}, bird-feather parameters are typically $\splus \approx 20$, $h/s \approx 0.5$ and $\ang \approx 30\degree$. Values for $\ngroove$ or $\wf$ were not given. Their SEM-pictures show that the feather barbs are comparable to blades with finite thickness and rounded tips. 
Although the present study was inspired by the bird-feather texture, it did not attempt to exactly reproduce that texture. Instead, texture parameters were chosen to allow a close comparison with the parallel-riblet studies described in subsections \ref{sec:parallel_riblets} and \ref{sec:parallel_riblets_yaw}. Specifically, ridge spacing and maximum blade height were fixed at $\splus = 17$ and $h/s = 0.5$.
Mainly two parameters were varied, namely $\ang$ and $\ngroove$. The angle $\ang$ was either $15\degree$ or $165\degree$, which permits the direct comparison with the conventional riblet texture in 15 degrees yaw. Larger angles (e.g. $\ang = 30\degree$ or $\ang = 150\degree$) were not considered, as \autoref{fig:drag_parallel_riblets} suggests that such angles will result in a drag increase. The number of grooves $\ngroove$ was varied between 1 and 128, which resulted in a feather-width change from $\wf = 0.10$ to $12.9$ in outer units or $\wf\plus = 35$ to $4506$ in wall units.

}

Apart from texture parameters, the other simulation parameters were (almost) the same for all herringbone simulations, which facilitates a fair comparison. Specifically, \autoref{tab:simulation_parameters} shows that domain size, grid resolution, Reynolds number and simulation time do not change much among the simulations. In addition, these parameters are also close to the ones for parallel-riblet (without/in yaw) simulations. All herringbone simulations were performed at $\rebulk = 5500$. Based on the validation of blade-riblet simulations (see \autoref{sec:parallel_riblets}), a low-Reynolds-number effect can be anticipated. However, as all simulations were performed at the same $\rebulk$, a proper comparison can still be made. The number of grid cells per groove $\ncg = 16$, so 16 grid cells span the groove width.

To substantiate the accuracy of the used numerical methods, two issues have to be addressed.
The first issue relates to the boundary conditions at the texture surface. Appendix \ref{sec:BC} shows that the no-slip and no-penetration conditions are sufficiently satisfied.
The second issue relates to the flow around the blade tip. The exact solution is singular there, which might introduce errors in the numerical solution. The improvement of the numerical accuracy for problems involving singularities is a research in itself \citep{shi2004combined}, but it is not the focus of the current study. For the present purpose, the used IBM is sufficiently accurate. Very similar IBMs have been used in DNSs of flow around other obstacles with sharp corners, such as cubes \citep{breugem2005, orlandi2006dns}, square and triangular elements \citep{orlandi2006turbulent}, and a flat plate normal to the free stream \citep{saha2007far, narasimhamurthy2009numerical}. The penultimate example shows that the IBM can deal with obstacles that are not aligned with the Cartesian grid, whereas the last example demonstrates that the IBM can also accurately capture separating flows at sharp corners.

\begin{figure}[t!]
\centering
\begin{minipage}[t]{7.5cm} 
\centering
\includegraphics[width=\textwidth]{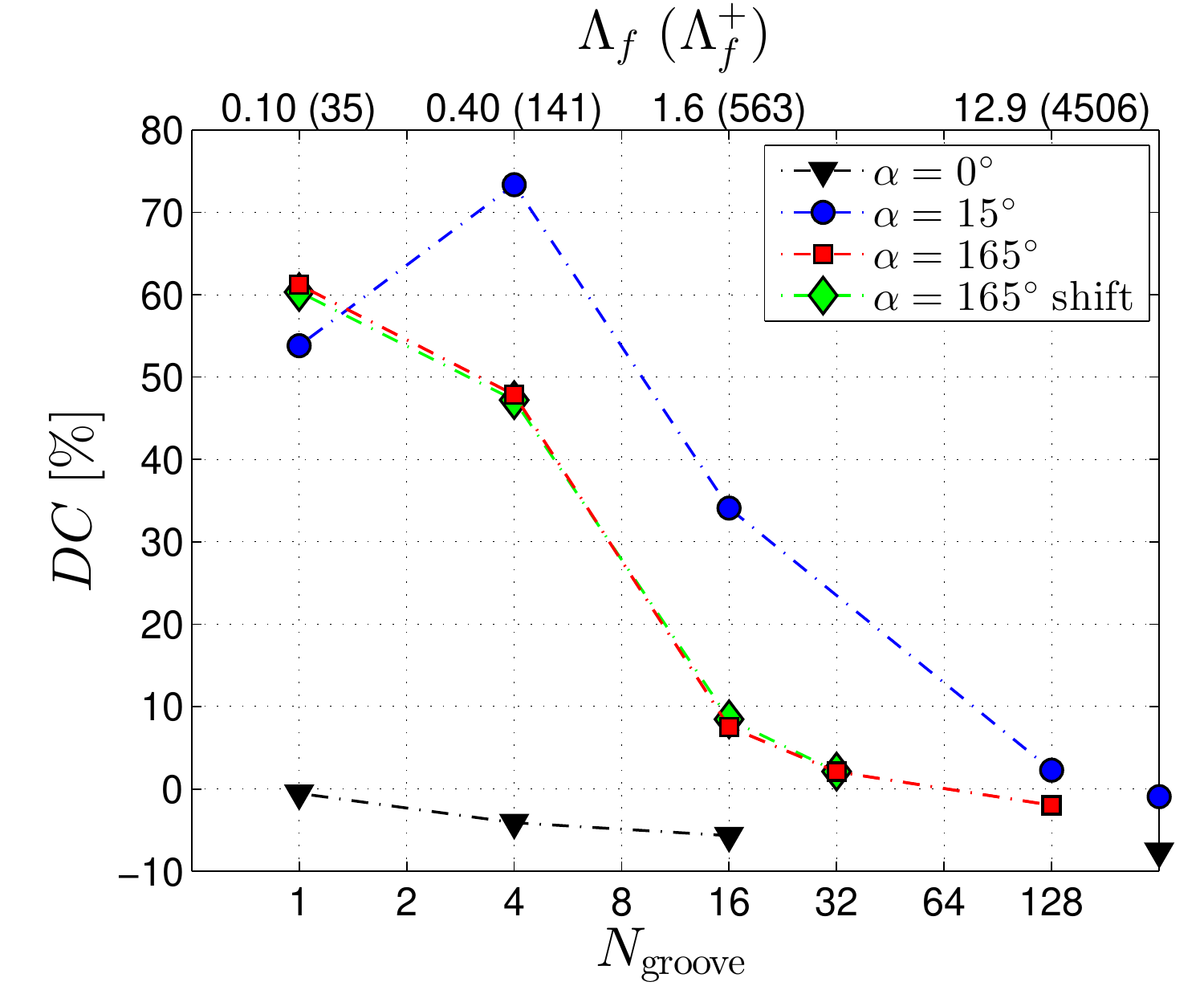}
\caption{Drag change as function of feather width for the herringbone riblet geometry, including results for the parallel ($\ang = 0\degree$) and shifted variants (see \autoref{fig:model_herringbone_texture_variations}).
The shown values of $\wf$ and $\wf^{+}$ apply to the textures with $\ang = 15\degree$ or $165\degree$, and are approximate for $\ang = 0\degree$ textures.
The data points on the right vertical axis belong to conventional parallel riblets with yaw angle $\ang = 0\degree$ and $\ang = 15\degree$.
}
\label{fig:DC_vs_ngroove}
\end{minipage}
\end{figure}

\subsection{Drag Reduction}

\textbf{Trend}
\autoref{fig:DC_vs_ngroove} shows the drag change as function of feather width for the herringbone riblet geometry. Drag changes significantly with the spanwise texture wavelength  or feather width. For the texture with $\ang = 165\degree$, the drag increases by 61\% for very narrow feathers, while the drag decreases by 2\% for very wide feathers. In general, wider feathers experience less drag. Only the texture with $\ang = 15\degree$ and $\ngroove = 1$ does not comply with this trend.
Drag reduction was only obtained when the texture approaches the parallel riblet texture in yaw, i.e. in the limit of very wide feathers. Only 2\% reduction was found in that limit, which is indeed comparable to the 0.9\% reduction that was obtained for parallel riblets at yaw angle $\ang = 15\degree$. It is significantly less than the 7.6\% reduction for parallel riblets aligned with the mean flow.



\textbf{Difference forward and backward flow}
There is a clear drag difference between textures with $\ang = 15\degree$ (forward flow) and $\ang = 165\degree$ (backward flow). For example, at $\ngroove = 16$ drag increases by 34\% for $\ang = 15\degree$, and only by 7.5\% for $\ang = 165\degree$. This shows that mean-flow reversal has a considerable effect on drag, in agreement with \citet{chen2014flow}.
Here, forward flow experiences more drag than backward flow, except for the $\ngroove = 1$ texture. 

The drag difference between forward and backward flow decreases for increasingly wide feathers, as can be expected. The textures with $\ang = 15\degree$ and $165\degree$ differ only because of riblet convergence or divergence in the feather shaft or feather edge regions. When $\wf \rightarrow \infty$, only a very small portion of the complete texture consists of converging or diverging riblets, so their contribution to the total drag becomes negligible.




\textbf{Standard texture variations}
\autoref{fig:DC_vs_ngroove} also shows the drag for the shifted and parallel riblet ($\ang = 0\degree$) variations. The drag of textures with and without shift is about the same for all feather widths. Hence, shifting of the top wall texture has almost no effect on drag, although the next subsection will explain that the mean flow is different for some cases.

In contrast with most herringbone textures, the parallel riblet texture with spanwise riblet height variation is able to reduce drag. The drag reduction is the least for $\ngroove = 1$, namely 0.5\%. The texture with $\ngroove = 1$ and $\ang = 0\degree$ is the conventional parallel-riblet geometry with $\splus = 17$, but with every second blade removed. It is known that this geometry is not optimal for drag reduction. When $\ngroove$ increases, drag reduction increases as well. In the limit of very large $\ngroove$, the texture approaches the standard parallel-riblet geometry, apart from a very slight spanwise height variation. Therefore, in that limit one might expect the drag reduction to be close to that for standard riblets. \newline







\begin{figure}[t!]
\centering
\begin{minipage}[t]{1\linewidth}
\centering
\includegraphics[width=\textwidth]{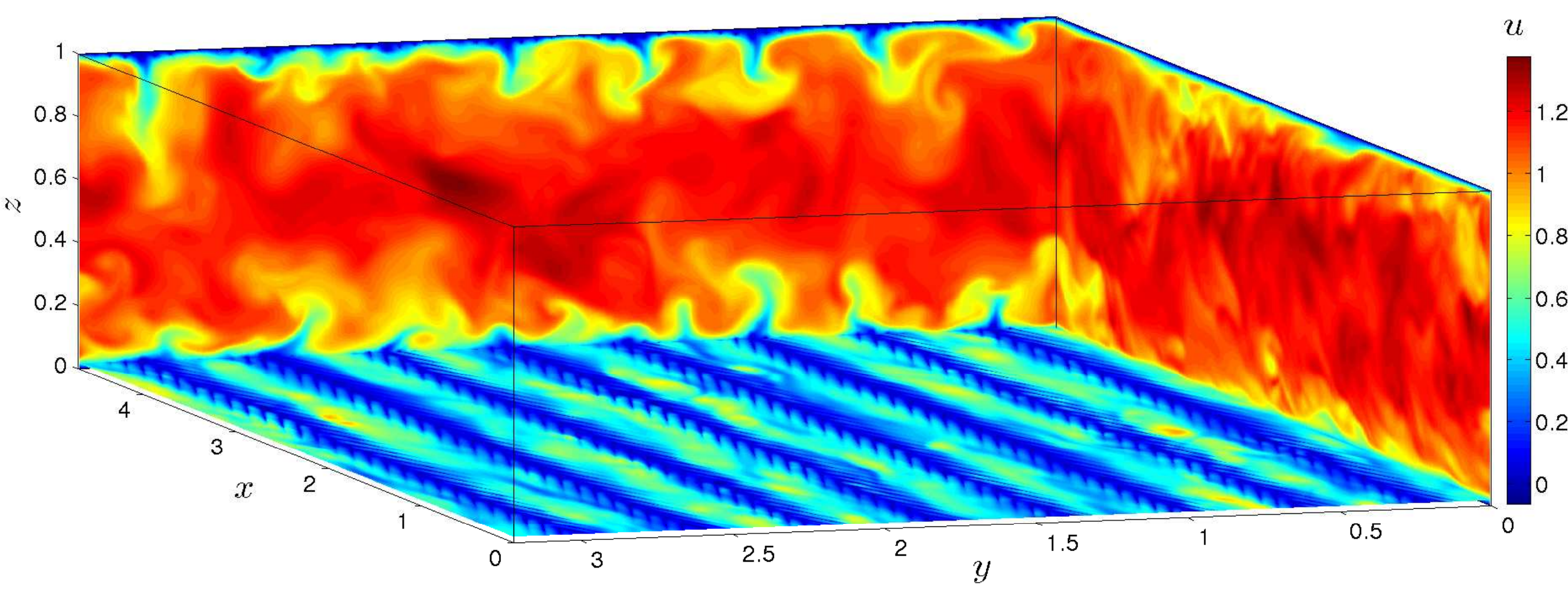}
\end{minipage}
\caption{Three cross sections of the instantaneous streamwise velocity in the simulation domain that belongs to the herringbone texture with $\ngroove = 4, \ang = 165\degree$. The horizontal plane is located at $z = 0.016$ or $z/h = 0.68$.}
\label{fig:instantaneous_flow}
\end{figure}

\begin{figure*}[t!]
\begin{minipage}[t]{1\linewidth}
\centering
\includegraphics[width=\textwidth]{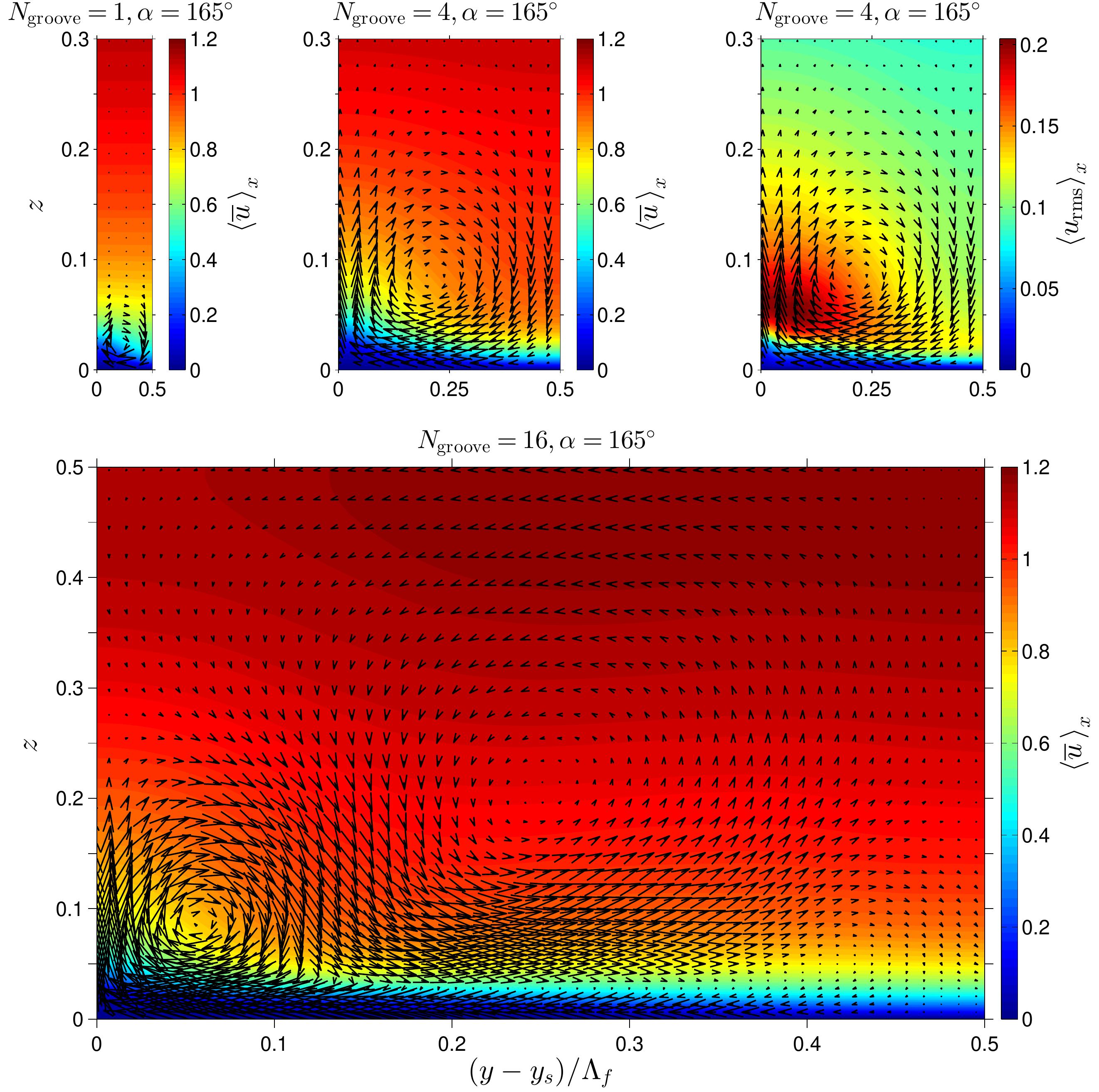}
\end{minipage}
\caption{Streamwise-averaged mean flow in a plane perpendicular to the streamwise direction for herringbone textures with $\alpha = 165\degree$ and $\ngroove = 1$ (\textbf{top left}), $4$ (\textbf{top center and right}), $16$ (\textbf{bottom}). Vectors exhibit in-plane secondary flow. Contours represent streamwise velocity (\textbf{top left}, \textbf{top center}, \textbf{bottom}) or streamwise velocity fluctuations (\textbf{top right}, $u_\mathrm{rms} = ( \tav{u^2} - \tav{u}^2 )^{1/2}$).}
\label{fig:mean_secondary_flow_herringbone}
\end{figure*}

\subsection{Flow Description}

\textbf{Instantaneous streamwise velocity}
To understand the drag-reduction results presented in the previous subsection, a detailed flow analysis is indispensable. 
\autoref{fig:instantaneous_flow} shows three cross sections of the instantaneous streamwise velocity. The V-shaped contours in the horizontal cross section reveal the presence of the herringbone texture. The plumes that appear in the $yz$-plane represent up- or downdrafts that result from the converging/diverging riblets in the texture.


\autoref{fig:mean_secondary_flow_herringbone} shows part of the mean streamwise-averaged flow fields for textures with $\ang = 165\degree$ and $\ngroove = 1, 4, 16$. Because of flow symmetry, the figures show only one feather half. The feather shaft is located at the left side and the feather edge at the right side of the figures.

$\boldsymbol{\alpha = 165\degree}$
For textures with $\ang = 165\degree$, the flow near the shaft converges, a local updraft of fluid results, and an approximately square vortex appears. The vortex does not reach the channel centerline, probably because of the small misalignment of the riblets with the mean flow ($\ang = 165\degree$ is relatively close to $180\degree$). For $\ngroove \geq 4$, it is limited in vertical extent to approximately $z = 0.2$, as can be seen for $\ngroove = 4$ and $16$ in the figure.
Because of spanwise confinement, for small feather widths the vortex decreases in size. For $\ngroove = 1$, it only reaches $z = 0.05$. Secondary flow is thus suppressed for small spanwise roughness spacings. This accords with the experimental finding that secondary flow disappears for decreasing spanwise spacing of roughness elements \citep{vanderwel2015effects}.

While mainly one vortex constitutes the secondary flow for $\ngroove = 1$ and $4$, tertiary flows appear for larger feather widths. For example, the mean flow for $\ngroove = 16$ (see \autoref{fig:mean_secondary_flow_herringbone}) shows a counterclockwise-rotating flow that extends to the channel centerline. Less pronounced is the small clockwise-rotating vortex near the feather edge. Several tertiary flows were also observed for even wider feathers ($\ngroove = 32, 128$). This agrees with the experimental finding that tertiary flows appear when the spanwise spacing of roughness elements increases above the boundary layer thickness \citep{vanderwel2015effects}.
These tertiary flows are likely similar to the secondary flows that form over streamwise-aligned roughness strips \citep{anderson2015numerical}. The latter are stress-induced (i.e. Prandtl’s secondary flows of the second kind), as opposed to the flow-curvature-induced secondary flows (i.e. Prandtl’s secondary flows of the first kind).

$\boldsymbol{\alpha = 15\degree}$
Reversing the flow direction from backward to forward also changes the secondary flow direction. For textures with $\ang = 15\degree$, riblets near the shaft diverge, resulting in a local downdraft of fluid and the appearance of a counterclockwise-rotating secondary flow to the right of the shaft. A tertiary flow again appears for $\ngroove \geq 16$. However, this time it is less well-defined, as its rotation direction is also counterclockwise. So, the most dominant tertiary flow does not change rotation direction by flow reversal.

\begin{figure}[t!]
\centering
\begin{minipage}[t]{0.93\linewidth}
\centering
\includegraphics[width=\textwidth]{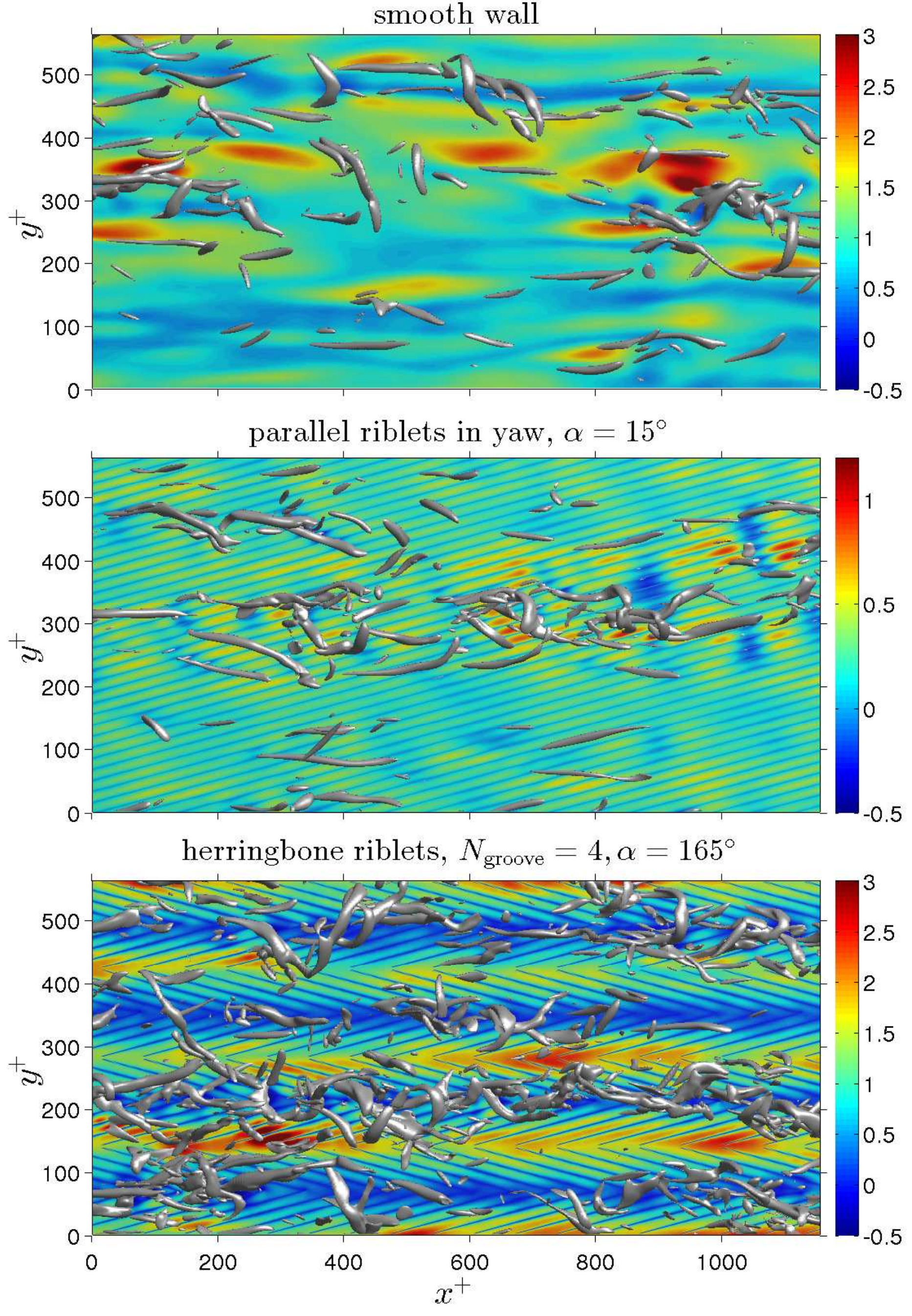}
\end{minipage}
\caption{
\change{
Instantaneous vortical structures and wall shear in a part of the bottom channel half. Vortical structures are iso-surfaces of the second invariant of the velocity gradient tensor with $Q^{+} = 0.03$. The colored contours represent the wall shear stress $\tau_w^{+}$ on the bottom wall. Three cases are shown (from top to bottom): smooth walls; parallel riblets in yaw of $\ang = 15\degree$; herringbone riblets with $\ngroove = 4, \ang = 165\degree$. $\rebulk = 5500$ for all cases and $\splus = 17$ for all textures.
}}
\label{fig:instantaneous_vortical_structures}
\end{figure}

\textbf{Shifted variant}
Shifting of the top-wall textures changes the mean-flow patterns only slightly. For all cases, the dominant vortex near the top wall shifts in the spanwise direction. As this vortex does not reach the channel centerline, it has no noticeable effect on the mean flow in the bottom channel half.
For $\ngroove = 1$ and $4$, this spanwise shift is the only change of the mean flow. 
For $\ngroove = 16$, however, the pronounced counterclockwise-rotating tertiary flow is also modified. It reaches the centerline for the not-shifted texture (see \autoref{fig:mean_secondary_flow_herringbone}), but it occupies the whole channel for the shifted variant. It is almost square, as for $\ngroove = 16$ the feather half-width ($\wf/2 = 0.80$) approximately equals the channel height ($\lz = 1$). Tertiary flow for $\ngroove = 32$ also extends from bottom to top wall. However, these modified tertiary flows apparently have not much influence on drag (considering \autoref{fig:DC_vs_ngroove}).

\textbf{Spanwise modulation}
The strong secondary-flow vortex near the shaft causes a spanwise modulation of the boundary layer, as is clear from the streamwise-velocity contours in \autoref{fig:mean_secondary_flow_herringbone}. Regions of updrafts (downdrafts) are characterized by low (high) streamwise velocity. The trend for streamwise velocity fluctuations is opposite, as the top-right subfigure shows: updrafts (downdrafts) are associated with increased (decreased) fluctuations. The same was found in experimental studies of convergent/divergent riblets \citep{koeltzsch2002, nugroho2013large} and in DNSs of turbulent boundary layers with uniform blowing/suction \citep{kametani2011direct}.
%

\change{

\textbf{Instantaneous vortical structures and wall shear}
%
%
Figure \ref{fig:instantaneous_vortical_structures} shows the instantaneous vortical structures and wall shear for three cases, namely smooth walls, parallel riblets in yaw and herringbone riblets. The vortical structures are iso-surfaces of the second invariant of the velocity gradient tensor (the $Q$-criterion, see e.g. \cite{dubief2000coherent}). The contours represent the shear stress $\tau_w^{+}$ on the bottom wall.

The smooth-wall plot shows some well-known features, such as low-speed streaks and hairpin-type vortical structures. The streaks are much less apparent in the second plot, which is attributed to the parallel riblets that adjust the turbulence. The streaks are more evident in the flow above the riblets. The vortices are comparable to the ones for the smooth wall. The plot for parallel riblets without yaw is not shown, as it is very similar to the one for parallel riblets in yaw.

The herringbone riblet texture exhibits the largest changes in vortical structures and shear. High shear is associated with diverging riblets, and low shear with converging riblets. The vortices seem to be ordered as well: they are abundant over regions with updrafts, but almost absent over regions with downdrafts. The same trend was again found for blowing/suction: vortices are enhanced by blowing in spite of the reduced wall shear stress, while vortices are suppressed by suction despite the increase of wall shear stress \cite{kametani2011direct}.

}

\subsection{Secondary Flow Strength} \label{sec:secondary_flow_strength}

\begin{figure*}[t!]
\begin{minipage}[t]{1\linewidth}
\centering
\includegraphics[width=\textwidth]{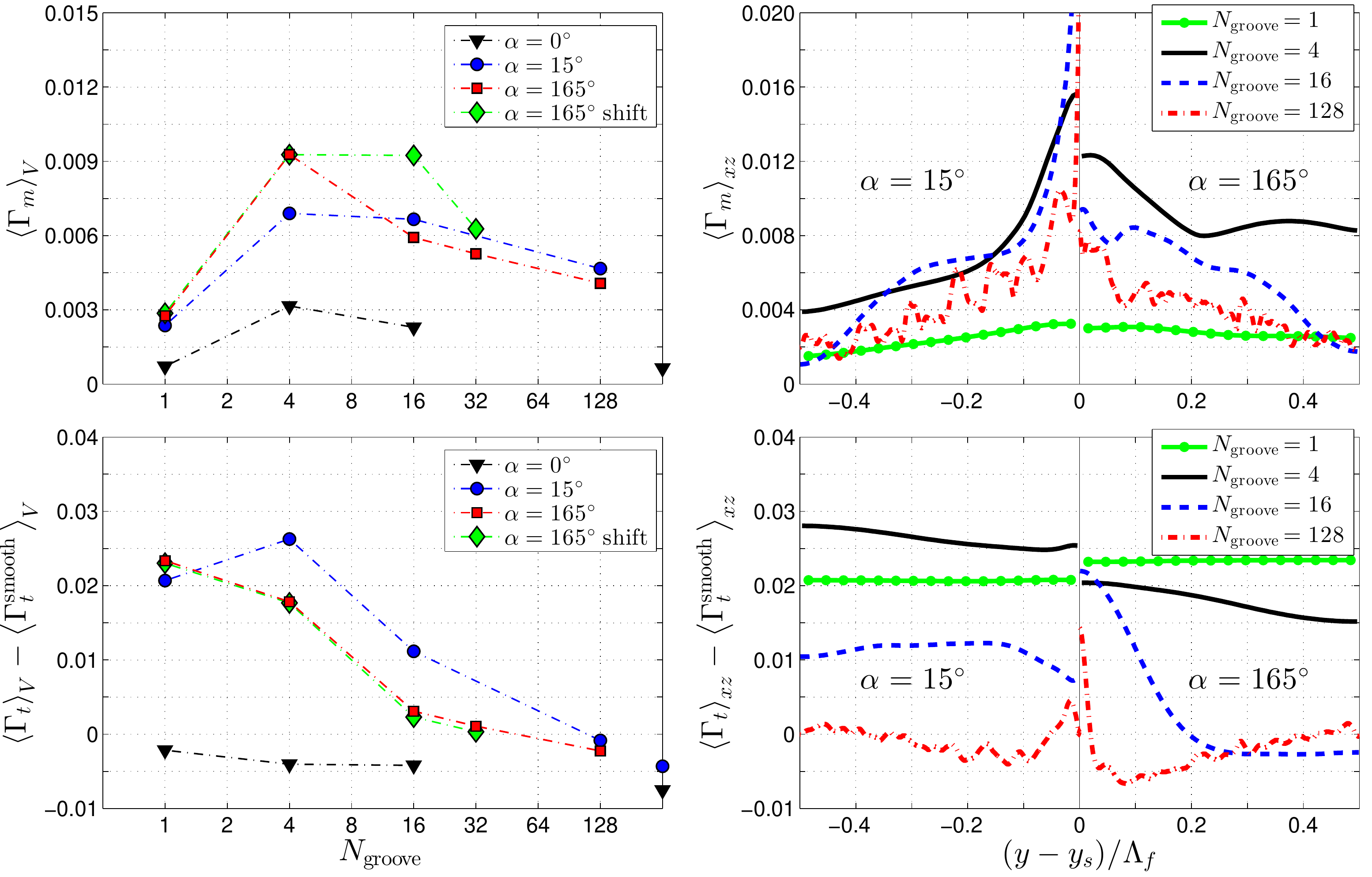}
\end{minipage}
\caption{Strength of mean and turbulent secondary flow for the herringbone riblet geometry.
\textbf{Top left:} Volume-averaged mean-secondary-flow-strength as function of feather width (specified by $\ngroove$). The data points on the right vertical axis represent the conventional parallel blade riblets with yaw angle $\ang = 0\degree$ and $\ang = 15\degree$ (not visible with $\vav{\secflowmean} = 0.027$).
\textbf{Top right:} Streamwise- and wall-normal-averaged mean-secondary-flow-strength as function of spanwise distance for textures with $\ang = 15\degree$ (left part) and $\ang = 165\degree$ (right part).
\textbf{Bottom left:} turbulent-secondary-flow-strength plotted as in top-left subfigure.
\textbf{Bottom right:} turbulent-secondary-flow-strength plotted as in top-right subfigure. In the bottom figures, the turbulent-secondary-flow-strength of the smooth wall $\vav{\secflowturb^\mathrm{smooth}} = \avxz{\secflowturb^\mathrm{smooth}} = 0.069$ is subtracted.
}
\label{fig:secondary_flow_strength_vs_ngroove}
\end{figure*}

To quantify the strengths of the secondary flows described in the previous subsection, the secondary flow strength $\secflow$ is introduced:
\begin{equation}
\begin{aligned}
\secflow 		& = \sqrt{\tav{v^2 + w^2}} = \sqrt{\secflowmean^2 + \secflowturb^2}, \\
\secflowmean	& = \sqrt{ \tav{v}^2 + \tav{w}^2 }, \\
\secflowturb	& = \sqrt{ \tav{(v')^2} + \tav{(w')^2} },
\end{aligned}
\label{eq:secondary_flow_strength}
\end{equation}
with $v' = v - \tav{v}$ and $w' = w - \tav{w}$. The secondary flow strength is decomposed into the mean-secondary-flow strength $\secflowmean$ and the turbulent-secondary-flow strength $\secflowturb$. These variables still depend on the spatial coordinates.
A volume- and $xz$-average of $\secflowmean$ and $\secflowturb$ is shown in \autoref{fig:secondary_flow_strength_vs_ngroove}.


\textbf{Influence of feather width}
\autoref{fig:secondary_flow_strength_vs_ngroove} (left part) shows that both mean and turbulent secondary flow become stronger with decreasing feather width. This is attributed to the converging or diverging riblets that trigger updrafts and downdrafts. When the feather width decreases, the spanwise density of these texture-generated secondary flows increases. In other words, the secondary flows that are created near the feather shaft and edges cover a relatively larger portion of the total fluid volume. This yields a stronger volume-averaged secondary flow for smaller feather widths.

\textbf{Mean and turbulent contribution}
The mean- and turbulent-secondary-flow strengths follow the same trend for most textures (namely an increase with a decrease of $\wf$). For all $\ngroove$, the converging or diverging riblets near the shaft trigger updrafts and downdrafts. As \autoref{fig:instantaneous_flow} demonstrates, these are not steady flow patterns. Instead, they can be considered as fluctuating ejections and sweeps that are generated by the texture. The resulting instantaneous secondary flow ($v^2 + w^2$) projects both onto $\secflowmean$ and $\secflowturb$ (see \autoref{eq:secondary_flow_strength}). A stronger instantaneous secondary flow yields in general an increase of both the mean and turbulent secondary flow. That explains why the mean and turbulent strength follow the same trend for $\ngroove \geq 4$.
The results for $\ngroove = 1$, however, deviate in this respect: $\vav{\secflowmean}$ more than halves as compared to $\ngroove = 4$. This suppression of mean secondary flow is due to spanwise confinement (see previous subsection). However, the fluctuating updrafts and downdrafts are not suppressed, so the turbulent secondary flow remains strong.

\textbf{Change with spanwise distance}
\autoref{fig:secondary_flow_strength_vs_ngroove} (top right) shows how $\secflowmean$ varies with spanwise distance.
The mean secondary flow is clearly strongest near the shaft, which is due to the counter-rotating vortices that form there. Such vortices are also generated near the feather edges, but the riblets have a small height there, which yields only a relatively weak secondary flow.
Compared to $\ang = 165\degree$, textures with $\ang = 15\degree$ have a stronger mean secondary flow at the shaft, which is probably due to the downdraft of high-momentum fluid there.

\autoref{fig:secondary_flow_strength_vs_ngroove} (bottom right) shows how $\secflowturb$ varies with spanwise distance.
The relatively uniform turbulence for the textures with $\ngroove = 1, 4$ shows that the flow is well-mixed. In contrast, the turbulence changes with $y$ for the other textures. When approaching the shaft, turbulence reduces when $\ang = 15\degree$ and increases when $\ang = 165\degree$. The same was observed for the streamwise velocity fluctuations in the previous subsection. When compared with the smooth wall, however, the turbulence at the shaft is increased for all textures. Away from the shaft, three textures exhibit local turbulence reductions.

\textbf{Variants}
\autoref{fig:secondary_flow_strength_vs_ngroove} also shows the volume-averaged secondary flow strength for the shifted texture variation. The turbulent contribution is practically the same as for the not-shifted texture. The same applies to the mean contribution for $\ngroove = 1$ and $4$. This agrees with the observation that, apart from the spanwise shift of the main vortex near the top wall, the mean secondary flow does not change. For $\ngroove = 16$ and $32$, however, the mean secondary flow is stronger for the shifted texture, which is ascribed to the formation of a tertiary flow that extends from bottom to top wall.


The secondary flow for the texture variation with $\ang = 0\degree$ is much weaker than for the herringbone riblet geometries, which is attributed to alignment of the riblets with the mean flow. Both $\vav{\secflowmean}$ and $\vav{\secflowturb}$ are still larger than the values for the parallel riblet geometry with constant blade height, although they seem to approach those values in the limit of large $\ngroove$. The figure shows that parallel riblet geometries can reduce turbulence.

\subsection{Spanwise Transport of Streamwise Momentum} \label{sec:momentum_balance_spanwise}


\begin{figure*}[t!]
\centering
\begin{minipage}[t]{1\linewidth}
\centering
\includegraphics[width=\textwidth]{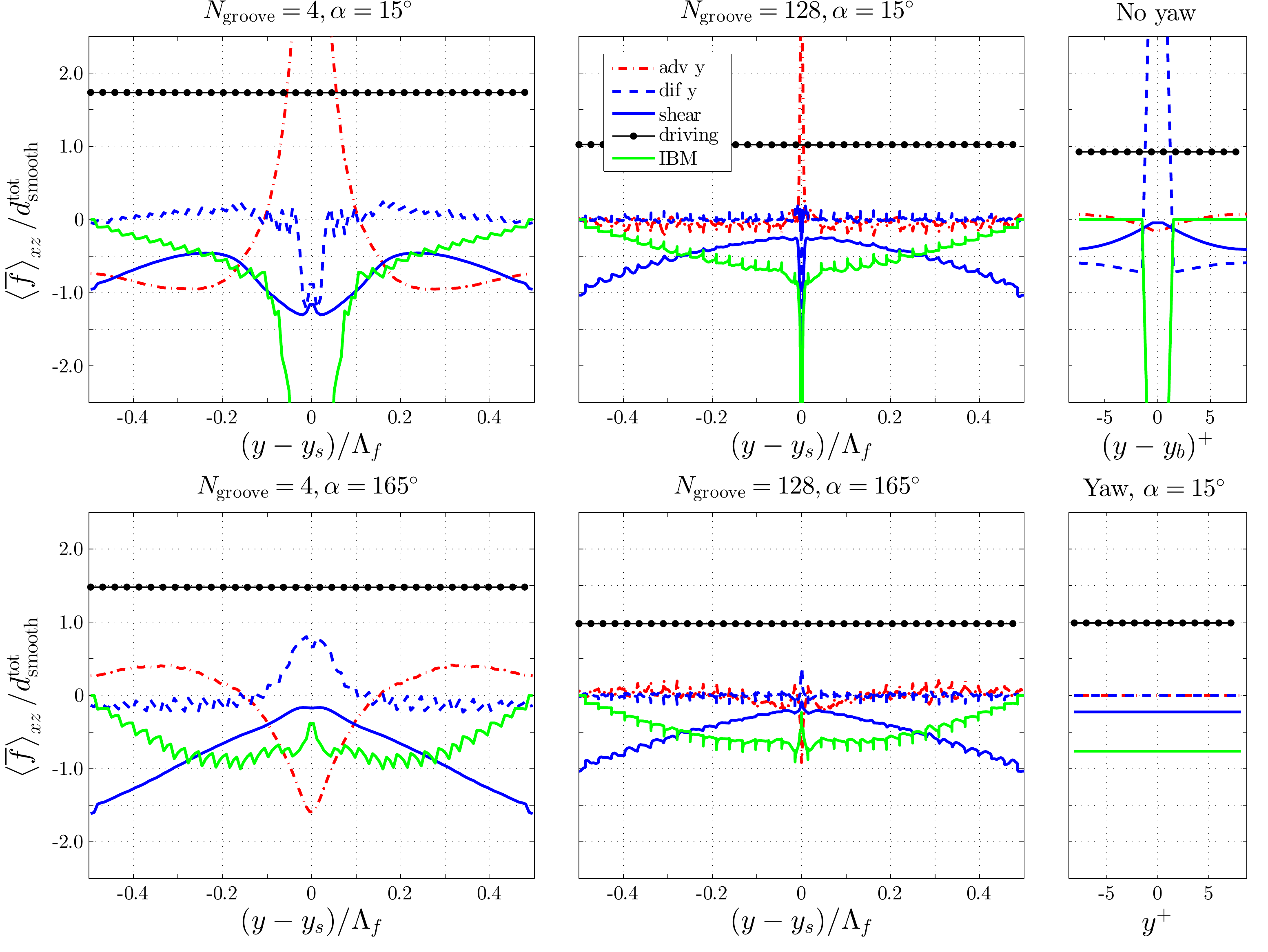}
\end{minipage}
\caption{Streamwise momentum balance as function of the spanwise coordinate (\autoref{eq:streamwise_momentum_balance_vs_y}). 
Six textures are shown:
narrow herringbone feathers ($\ngroove = 4$) for $\ang = 15\degree$ (\textbf{top left}) and $\ang = 165\degree$ (\textbf{bottom left}); 
wide herringbone feathers ($\ngroove = 128$) for $\ang = 15\degree$ (\textbf{top center}) and $\ang = 165\degree$ (\textbf{bottom center});
parallel riblets aligned with the mean flow (\textbf{top right}) and in yaw of $\ang = 15\degree$ (\textbf{bottom right}).
The blade location is denoted by $y_b$.
For all cases, $\rebulk = 5500$ and $\splus = 17$.}
\label{fig:mom_balance_vs_y}
\end{figure*}

To confirm that advective transport is most important around the shaft, this subsection considers the streamwise momentum balance as function of spanwise distance. The Navier-Stokes equation for streamwise momentum (see \autoref{eq:Navier_Stokes_equations}) is rewritten as $\partial u / \partial t = \sum f$, where $f$ is one of the terms in that equation. A time, streamwise and wall-normal average (denoted by $\avxz{\tav{f}}$) is applied to this equation. Assuming statistical stationarity and using the boundary conditions, the balance becomes:
\begin{align}
0 = & \underbrace{- \avxz{\pd{\tav{uv}}{y}}					}_{\text{adv y}}
	+ \underbrace{\avxz{\frac{1}{\rebulk}\pds{\tav{u}}{y}}	}_{\text{dif y}}
	+ \underbrace{\avxz{\frac{1}{\rebulk}\pds{\tav{u}}{z}}	}_{\text{shear}}
	+ \underbrace{\avxz{\gammaf \tav{\fdriving}}			}_{\text{driving}}
	+ \underbrace{\avxz{\tav{\fibmx}}						}_{\text{IBM}}.
\label{eq:streamwise_momentum_balance_vs_y}
\end{align}
The terms in this equation only depend on the spanwise coordinate $y$. The third term is called `shear', since it equals the sum of bottom and top-wall shear as a result of the wall-normal integration.
All terms are divided by $\ftotsmooth$ and shown in \autoref{fig:mom_balance_vs_y} for four herringbone textures. The abscissa represents the spanwise distance to the feather shaft. Most curves exhibit approximately equidistant wiggles that result from staircasing: the riblet height increases in 17 steps of size $\dzw$ from 0 at the feather edges to $h$ at the shaft.





%
Before the differences between $\ngroove = 4$ and $128$ are highlighted, the general behavior of the different terms is clarified.
The driving term is almost constant with $y$, because $\gammaf$ changes only marginally from the feather shaft to the edges. For $\ngroove = 4$, this term is clearly larger than 1, indicating a significant drag increase.

The shear term is negative, as it tends to decelerate the fluid. Its magnitude quantifies how the flat-wall streamwise shear changes with the spanwise coordinate.
When moving from the feather edges towards the shaft, the wall-shear term first decreases (in magnitude), which is attributed to shielding of the flat wall by blades which increase in height. This trend continues for textures with $\ang = 165\degree$. However, the shear magnitude peaks near the shaft for $\ang = 15\degree$, which is ascribed to the local downdraft that transports high momentum towards the wall. For very wide feathers ($\ngroove = 128$), the shear term equals -1 at the feather edges. There the blade height is zero and the smooth-wall result is recovered.


The IBM term represents the streamwise drag force on the riblets. It is negative, as it is responsible for a velocity decrease (like the shear term). It equals zero at the feather edges, because the riblets have no height there. When approaching the feather shaft from the edges, the IBM-force magnitude first increases, which is due to riblet-height increase. Near the shaft, it has a local minimum for $\ang = 165\degree$. In contrast, it is very large there for $\ang = 15\degree$, likely due to the downdraft of high-speed fluid.


The balance for $\ngroove = 4$ shows significant contributions from spanwise advective and diffusive transport, especially near the shaft. 
For $\ang = 15\degree$, advection is on average responsible for an increase of streamwise momentum near the shaft. This is attributed to the secondary-flow vortex that transports low momentum away from the shaft (near the wall) and high momentum towards the shaft (closer to the channel centerline). In contrast, diffusion transports high momentum away from the shaft. These trends are opposite to that of textures with $\ang = 165\degree$, for which spanwise advection causes a streamwise momentum decrease and diffusion an increase near the shaft.


Compared to $\ngroove = 4$, the relative importance of the terms in \autoref{eq:streamwise_momentum_balance_vs_y} is very different for $\ngroove = 128$. The prominent peaks near the shaft are very narrow. Spanwise transport by advection and diffusion is close to zero for the largest part of the feather. Away from the shaft, the texture behaves as parallel riblets in yaw with a local balance between the driving force on the one hand, and the IBM and wall-shear force on the other hand.
\subsection{Wall-normal Transport of Streamwise Momentum} \label{sec:momentum_balance_wall_normal}



\begin{figure*}[t!]
\centering
\begin{minipage}[t]{1\linewidth}
\centering
\includegraphics[width=\textwidth]{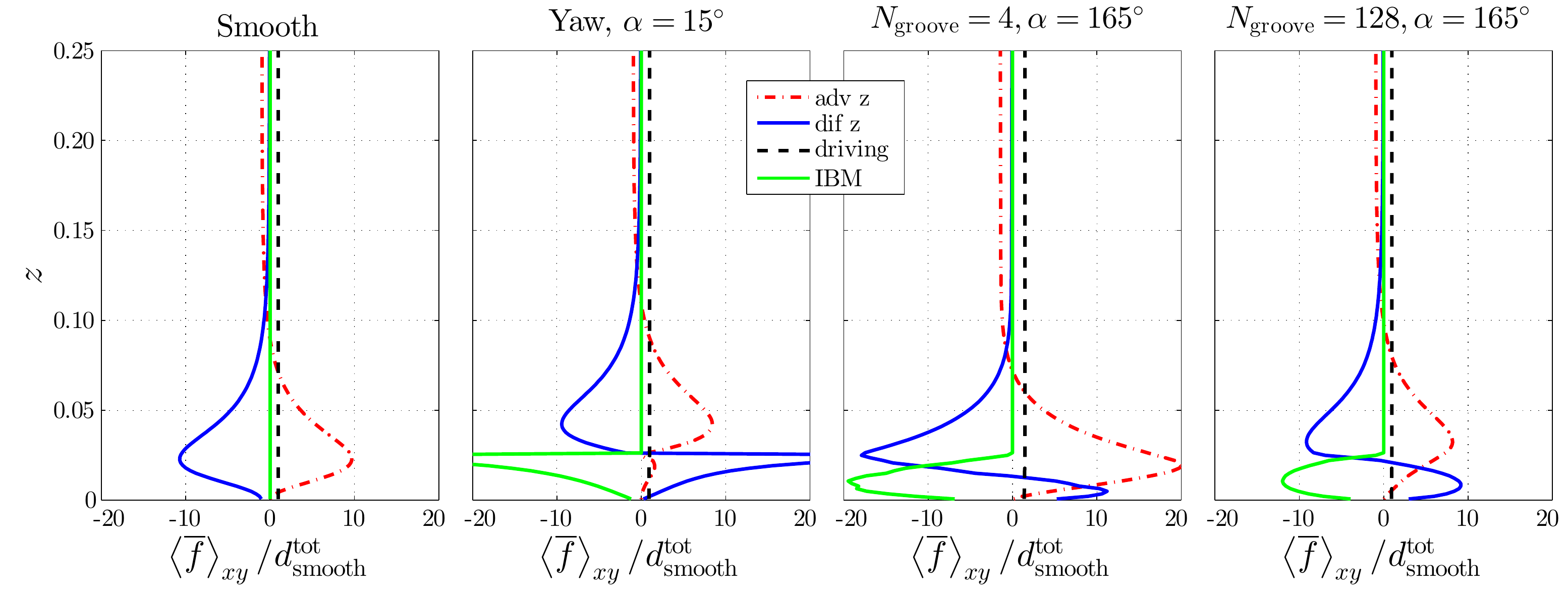}
\end{minipage}
\caption{Streamwise momentum balance as function of the wall-normal coordinate (\autoref{eq:streamwise_momentum_balance_vs_z}). Four cases are shown (from left to right):
smooth walls;
parallel riblets in yaw of $\ang = 15\degree$;
narrow herringbone feathers ($\ngroove = 4$) for $\ang = 165\degree$;
wide herringbone feathers ($\ngroove = 128$) for $\ang = 165\degree$.
$\rebulk = 5500$ for all cases and $\splus = 17$ for all textures.}
\label{fig:mom_balance_vs_z}
\end{figure*}

The previous subsections show that drag increase is accompanied by a strong secondary flow, which suggests that enhanced advection is responsible for the drag augmentation. To underpin this suggestion, this subsection considers the streamwise momentum balance as function of the wall-normal coordinate. A time, streamwise, and spanwise average (denoted by $\avxy{\tav{f}}$) is applied to the Navier-Stokes equation $\partial u / \partial t = \sum f$. Assuming statistical stationarity and using the boundary conditions, the balance reads:
\begin{align}
0 = \underbrace{- \avxy{\pd{\tav{uw}}{z}}					}_{\text{adv z}}
	+ \underbrace{\avxy{\frac{1}{\rebulk}\pds{\tav{u}}{z}}	}_{\text{dif z}}
	+ \underbrace{\avxy{\gammaf \tav{\fdriving}}			}_{\text{driving}}
	+ \underbrace{\avxy{\tav{\fibmx}}						}_{\text{IBM}}.
\label{eq:streamwise_momentum_balance_vs_z}
\end{align}
The terms in this equation only depend on the wall-normal coordinate $z$. All terms are divided by $\ftotsmooth$ and displayed in \autoref{fig:mom_balance_vs_z} for four cases. The profiles for parallel riblets without yaw (not shown) are very similar to that for riblets in yaw. In addition, the profiles for herringbone textures with $\ang = 15\degree$ (not shown) are similar to the ones for $\ang = 165\degree$. 
The behavior of the different terms is clarified below. The driving term is almost constant, as before.

The IBM term represents the drag force on the texture, so it is only present near the wall and it is responsible for a velocity decrease. The drag force is especially large near the blade tips. That explains the peak at blade height for riblets in yaw, for which all blades have the same height. The large IBM force near the blade tips is in \autoref{fig:mom_balance_vs_z} not evident for herringbone textures. Instead, the peak is smeared out due to the spanwise blade-height variation between $0$ and $h$ (see \autoref{eq:herringbone_riblet_height}).
For the two herringbone textures shown in the figure, the narrower feather clearly experiences a larger IBM force than the wider feather.

Diffusive transport is especially important near the wall. It has a negative tendency for smooth walls, but for textured walls it becomes positive in between the riblets. The latter is associated with an inflection point in the mean streamwise-velocity profile.

Advective transport is significant throughout the whole channel. Near the channel centerline, it balances the driving force (for all cases). Near the wall, turbulent advective transport is responsible for a velocity increase and the associated drag augmentation.
Parallel riblets suppress turbulent advection, as is clear from the second sub-figure. Compared to smooth walls, the peak of advection shifts upwards and shrinks. The herringbone case with $\ngroove = 128$ shows the same trend, although the upward shift is less pronounced. Turbulent transport below $z = h$ is less suppressed as compared to parallel riblets in yaw, which is ascribed to the blade height decrease with spanwise distance to the shaft. Still, weakening and lifting of advective transport is quite apparent.
In contrast, advective transport is much stronger for $\ngroove = 4$ as compared to a smooth wall. This reinforces the suggestion that drag augmentation is caused by enhanced advection, which will be confirmed in the next section with a quantitative analysis.

\section{Drag Change Decomposition} \label{sec:fik}


\subsection{Previous Research}


\citet{fukagata2002} derived an equation (called FIK-identity) that decomposes the frictional-drag coefficient into contributions from different dynamical effects. 
They distinguished four contributions, namely laminar (or bulk), turbulent, inhomogeneous and transient. For homogeneous and steady plane channel flow, the FIK-identity reads (in our notation):
\begin{align}
\frac{1}{12}\tav{\fdriving} = \frac{1}{\rebulk} + \intn{0}{1}{\left( \frac{1}{2}-z \right) \avxy{-\tav{u'w'}} }{z},
\label{eq:standard_FIK_identity}
\end{align}
where it is used that the domain height $\lz = 1$. Note that the skin-friction coefficient $C_f \equiv 2 \tav{\tauw^*}/\lb \rho^* \ub^{*2} \rb = \tav{\fdriving}$ for smooth-wall channel flow.

This relation has been extended to geometrically more complex surfaces by \citet{peet2009theoretical}. They derived analytical relations for streamwise-, spanwise- and quasistreamwise-homogeneous surfaces. To our knowledge, however, these relations cannot be used for the herringbone textures. That asks for a second extension of the FIK-identity.

\subsection{Derivation of Extended FIK-identity}

The extended FIK-identity is derived from the slab-averaged streamwise momentum balance (\autoref{eq:streamwise_momentum_balance_vs_z}). For ease of notation, that equation is written as $0 = \sum_{l} \left. f^{(l)} \right|_{z}$, where $f^{(l)}$ represents one of the terms in that equation. The subscript $z$ expresses that $f^{(l)}$ only depends on the wall-normal coordinate. In what follows, the global drag balance is needed. It follows from a single integration of \autoref{eq:streamwise_momentum_balance_vs_z}: 
$0 = \sum_{l} \intn{0}{1}{ \left. f^{(l)} \right|_{z} }{z}$ or
\begin{equation}
\begin{aligned}
 \dtot & = \vav{\gammaf \tav{\fdriving}} \\
 			   & = \avxy{\frac{1}{\rebulk} \left.\pd{\tav{u}}{z}\right|_{z = 0}}
 			       - \avxy{\frac{1}{\rebulk} \left.\pd{\tav{u}}{z}\right|_{z = 1}}
 			       - \vav{\tav{\fibmx}}.
\end{aligned}
\label{eq:global_drag_balance}
\end{equation}
Next, \autoref{eq:streamwise_momentum_balance_vs_z} is rewritten as 
$0 = \sum_{l} (1/2) \left( \left. f^{(l)} \right|_{z} + \left. f^{(l)} \right|_{1-z} \right)$ to explicitly account for symmetry in the mean flow. Triple integration is applied to this equation, such that the extended FIK-identity in condensed form reads:
\begin{align}
0 = \sum_{l}
\intn{0}{1}{\intn{0}{z}{\intn{0}{\ztilde}{ 
\frac{1}{2} \left( \left. f^{(l)} \right|_{\zhat} + \left. f^{(l)} \right|_{1-\zhat} \right)
 }{\zhat}}{\ztilde}}{z}.
\label{eq:}
\end{align}
Using the boundary conditions at the channel walls, the definition of the bulk velocity (i.e. $\intn{0}{1}{\sav{\tav{u}}_{xy}\!}{z} = 1$), integration by parts to transform multiple to single integrations, and the global drag balance (i.e. \autoref{eq:global_drag_balance}), the last equation becomes:
\begin{equation}
\begin{aligned}
\tav{\fdriving} \left[ \intn{0}{1}{ \left\lbrace \frac{1}{2} z \lb 1 - z \rb \right\rbrace \avxy{\gammaf} }{z} \right] & = \\
\frac{1}{\rebulk}
+ \intn{0}{1}{ \left\lbrace \frac{1}{2} - z \right\rbrace \avxy{-\tav{uw}} }{z}
& + \intn{0}{1}{ \left\lbrace \frac{1}{2} z \lb 1 - z \rb \right\rbrace \avxy{-\tav{\fibmx}} }{z}.
\end{aligned}
\label{eq:almost_extended_FIK_identity}
\end{equation}
%
%
%
%
To arrive at an equation for $\dtot = \langle \gammaf \tav{\fdriving} \, \rangle_{V} = \tav{\fdriving} \intn{0}{1}{\avxy{\gammaf}}{z}$, \autoref{eq:almost_extended_FIK_identity} is divided by the prefactor in square brackets and multiplied by $\intn{0}{1}{\avxy{\gammaf}}{z}$, which yields the final extended FIK-identity:
\begin{equation}
\begin{aligned}
\dtot 	& = \dbulk + \dmeanadv + \dturbadv + \dibm \\
		& = \sum_{l} d^{(l)}.
\end{aligned}
\label{eq:extended_FIK_identity}
\end{equation}
Notice that the total advection term is split up into contributions from mean and turbulent flow, using that $\tav{uw} = \tav{u}\,\tav{w} + \tav{u'w'}$.
The symmetry or antisymmetry with respect to the channel centerline at $z = 1/2$ is evident from the factors in braces in \autoref{eq:almost_extended_FIK_identity}.
As the multiplier of $\tav{\fdriving}$ in that equation depends on $\gammaf$, the bulk term $\dbulk$ is slightly texture dependent, as was also found by \citet{peet2009theoretical}. For smooth walls, $\gammaf = 1$ and that multiplier equals 1/12, such that \autoref{eq:standard_FIK_identity} is recovered.

As \autoref{eq:extended_FIK_identity} applies to both smooth and textured walls, the drag change can be decomposed in a similar way:
\begin{equation}
\begin{aligned}
\dc & = \frac{\ftot - \ftotsmooth}{\ftotsmooth} \\
   & = \sum_{l} \left\lbrace \frac{ d^{(l)} - \fsmooth^{(l)}}{\ftotsmooth} \right\rbrace 
   \equiv \sum_{l} dc^{(l)}.
\end{aligned}
\label{eq:DC_FIK_decomposition}
\end{equation}
The term within braces is abbreviated as $dc^{(l)}$. It represents the change of a certain term for textured walls as compared to that term for smooth walls. This decomposition quantifies which terms contribute to drag reduction or increase.


\change{

The identity presented here is somewhat different from the one derived by \citet{peet2009theoretical}. First, their decomposition only applies to skin friction, whereas the drag decomposition in \autoref{eq:extended_FIK_identity} also includes the pressure drag. Second, the IBM-term is not present in their identity. They used a body-fitted coordinate system in their derivation. As a result, the shear stress on the texture directly derives from integration of the viscous diffusion term. That approach has the added advantage that the skin-friction coefficient for simple textures in a laminar flow can be computed exactly based on purely geometrical considerations without performing the flow calculations \citep{peet2009theoretical}. However, their relation applies to quasi-homogeneous surfaces only and adopts a more complicated integration using a body-fitted grid. In contrast, the Cartesian integration that is employed here is not restricted to certain geometries.

The extended FIK-identity (\autoref{eq:extended_FIK_identity}) is not only useful when an Immersed Boundary Method (IBM) is used. In the present work, $\fibm_i$ is a body force that models the shear and pressure forces that the texture exerts on the flow. However, the FIK-identity applies to any body force. Furthermore, the current drag decomposition is also applicable to body-fitted calculations. In that case, the obstacles should be considered as part of the domain, because the identity is based on integration over the entire rectangular channel volume. Three steps are required for a successful use of the decomposition in this situation. 
1) The geometry should be translated into a three-dimensional phase-indicator function $\gammaf$. 
2) A zero-flow condition should be used for the obstacle volume. 
3) The drag force on the obstacle surface should be translated into a three-dimensional body force or IBM force.

}

In view of the factors between braces in \autoref{eq:almost_extended_FIK_identity}, advection and the IBM force contribute differently to the total drag. The weighing factor for advection is largest near the wall, so significant advective transport near the wall contributes most to drag augmentation. In contrast, the weighing factor for the IBM force is largest near the channel centerline. Obstructing the flow there is for two reasons more detrimental than an obstruction near the wall: drag increases due to a larger flow velocity (so a larger IBM force) and a larger weighing factor.




\tabcolsep=0.10cm
\begin{table*}[t!]
  \centering
  \begin{minipage}[t]{1\linewidth}
  \centering
  \caption{Contributions to the total skin friction for smooth-wall turbulent channel flow. The terms are as given in \autoref{eq:extended_FIK_identity} and divided by $\dtot$. The rest term equalizes the left-hand and right-hand side of that equation.}
    \begin{tabular}{ l l l l l l l }
    \hline
	$\rebulk$	 & $\dtot \, [\%]$	 & $\dbulk \, [\%]$	 & $\dmeanadv \, [\%]$	 & $\dturbadv \, [\%]$	 & $\dibm \, [\%]$	 & $\drest \, [\%]$ \\ 
\hline
	5500	 & 100	 & 26.9	 & $4.3\e{-17}$	 & 73.1	 & 0	 & -0.013 \\ 
	11000	 & 100	 & 16.0	 & $1.7\e{-17}$	 & 83.9	 & 0	 & 0.086 \\ 
	22000	 & 100	 & 9.6	 & $1.4\e{-16}$	 & 90.3	 & 0	 & 0.10 \\ 
\hline
    \end{tabular}
    \label{tab:FIK_smooth}
  \end{minipage}
\end{table*}

\subsection{Smooth Walls}

\autoref{tab:FIK_smooth} shows the decomposition of the total drag into the different contributions for smooth-wall turbulent channel flow. The results for three Reynolds numbers are shown. The total drag is taken as a reference. The rest term equalizes the left-hand and right-hand side of \autoref{eq:extended_FIK_identity}. It results from the finite simulation time. It becomes slightly larger for higher $\rebulk$, which is attributed to a relatively shorter simulation time. Its magnitude is typical for all the other simulations.

As the table shows, only bulk transport and turbulent advection contribute to the total drag in smooth-wall channel flows. The mean advection term is zero to machine precision, and $\dibm = 0$. At $\rebulk = 5500$, the total drag comes for 27\% from the bulk and for 73\% from the turbulent advection term. These numbers are close to the 26\% and 74\% reported previously for $\rebulk = 5460$ \citep{peet2009theoretical}. As the Reynolds number increases, the relative contribution of turbulence to the total drag increases. This stresses once more the need for reduction of turbulent drag in high-Reynolds-number flows.



\subsection{Parallel Riblets}

\begin{figure*}[t!]
\begin{minipage}[t]{1\linewidth}
\centering
\includegraphics[width=\textwidth]{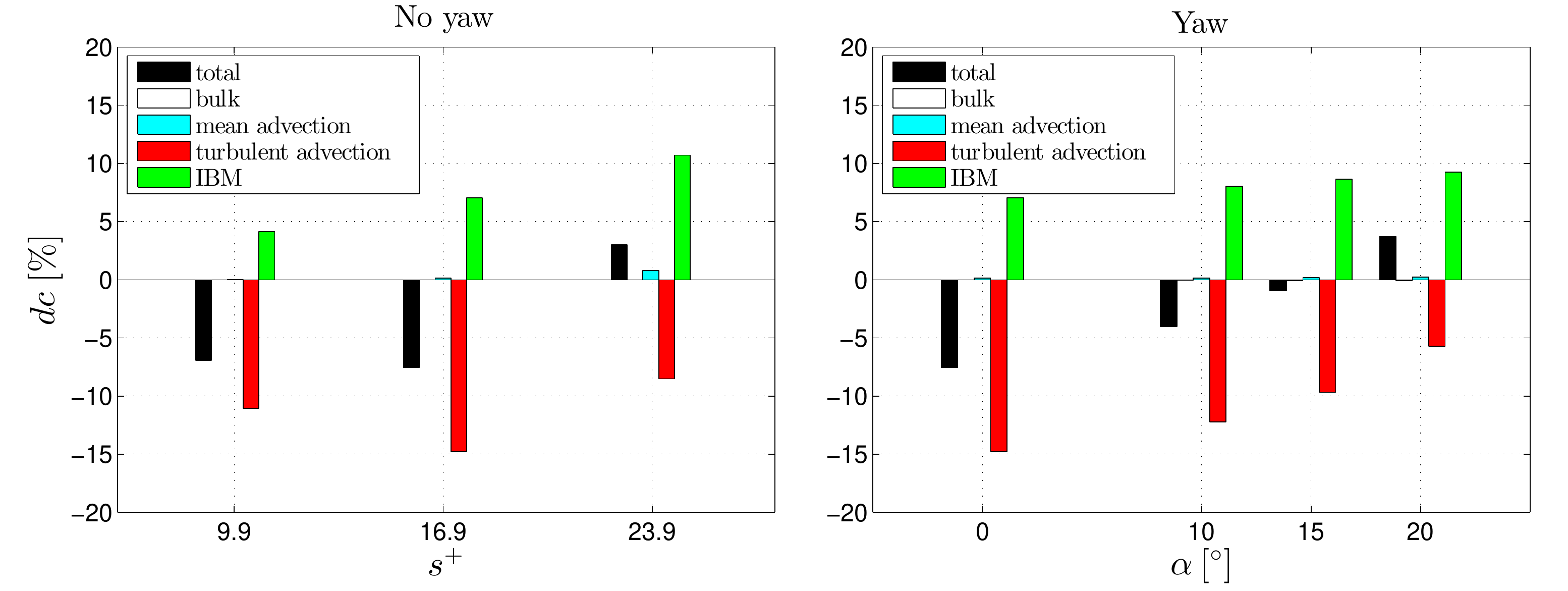}
\end{minipage}
\caption{Different contributions to the total drag change (\autoref{eq:DC_FIK_decomposition}) for parallel blade riblets. Note that five bars belong to only one abscissa.
\textbf{Left:} Decomposition as function of riblet spacing in wall units for riblets aligned with the mean flow ($\ang = 0\degree$).
\textbf{Right:} Decomposition as function of yaw angle for fixed riblet spacing $\splus = 17$.}
\label{fig:FIK_parallel}
\end{figure*}

Using \autoref{eq:DC_FIK_decomposition}, the drag change for parallel riblets is decomposed into contributions from the bulk, mean advection, turbulent advection and IBM terms. \autoref{fig:FIK_parallel} exhibits this decomposition as function of $\splus$ and $\ang$ in a bar graph. The total term equals the total drag change, which was shown already in \autoref{fig:drag_parallel_riblets}. The bulk term is in both graphs not visible, as it is very close to zero.

\textbf{Change with blade spacing}
The IBM term, which represents the drag force on the blades, is responsible for an increase of the total drag. Apart from being positive, it also increases with $\splus$. When the riblet spacing increases, the riblet height increases as well ($h/s = 0.5$ is fixed), but the total blade area per unit spanwise width remains constant. Therefore, the increase of the IBM term with $\splus$ cannot be due to an increased wetted area, but it is ascribed to blades that further protrude into the flow.
Although it remains small as compared to the other terms, the contribution from mean advection also increases with blade spacing. That is attributed to a stronger mean secondary flow for larger $\splus$ (not shown here, see also e.g. \citep{choi1993DNS, goldstein1998secondary}).
Finally, the turbulent drag contribution is negative, which indicates suppression of turbulent streamwise momentum transport. For the three $\splus$ values shown here, the maximum turbulent-drag suppression is almost 15\% at $\splus = 17$. The figure demonstrates that the optimum $\splus$ is a trade-off between an additional drag force on the blades and reduced turbulent transport because of the blades.

\textbf{Change with yaw angle}
\autoref{fig:FIK_parallel} shows a similar decomposition for parallel riblets in yaw at fixed $\splus = 17$.
The IBM term increases for increasing yaw angle $\ang$, which might be due to the additional pressure drag. 
The change in the mean advection term is very small.
The contribution from turbulent advection is increasingly less negative when $\ang$ increases. The figure thus indicates that deterioration of riblets in yaw is both due to an increased drag force on the blades and reduced suppression of turbulent transport.


\begin{figure*}[t!]
\begin{minipage}[t]{1\linewidth}
\centering
\includegraphics[width=\textwidth]{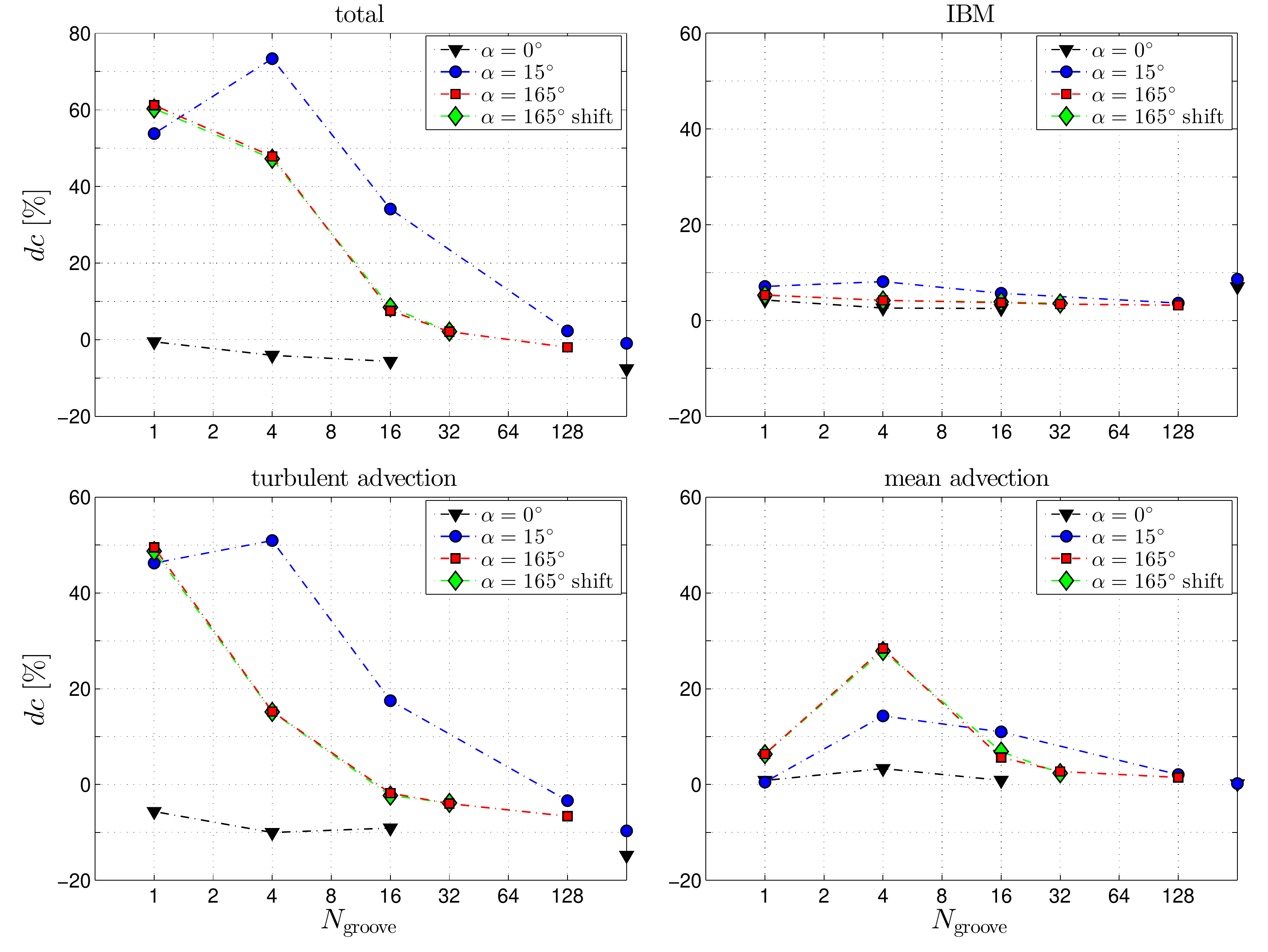}
\end{minipage}
\caption{Decomposition of the total drag change (\textbf{top left}) for herringbone riblets in contributions from the IBM-force (\textbf{top right}), turbulent advection (\textbf{bottom left}) and mean advection (\textbf{bottom right}) according to \autoref{eq:DC_FIK_decomposition}. All figures show the contribution change relative to a smooth wall as function of feather width. 
The data points on the right vertical axes represent the conventional parallel blade riblets with yaw angle $\ang = 0\degree$ and $\ang = 15\degree$.}
\label{fig:FIK_herringbone}
\end{figure*}

\subsection{Herringbone Riblets}

\textbf{Drag decomposition}
\autoref{fig:FIK_herringbone} shows the decomposition of the total drag change into its contributions from the IBM force, turbulent advection and mean advection. The bulk and rest terms are not presented. They never exceed 0.04\% (bulk) and 0.08\% (rest) in magnitude. The top left graph is the same as \autoref{fig:DC_vs_ngroove} and is shown here again for ease of comparison. 
Apart from the bulk contribution, the IBM term is the least important term for most textures. In the worst case, it causes an 8.1\% increase of drag, which is comparable to results for standard parallel riblets (7.0\% for $\ang = 0\degree$, 8.6\% for $\ang = 15\degree$).
Advection by the mean flow can be a significant contribution to drag increase, up to 28\% for $\ngroove = 4, \ang = 165\degree$.
The turbulent advection term exhibits the largest changes: it varies between a 6\% reduction and a 50\% increase.

\begin{figure*}[t!]
\begin{minipage}[t]{1\linewidth}
\centering
\includegraphics[width=\textwidth]{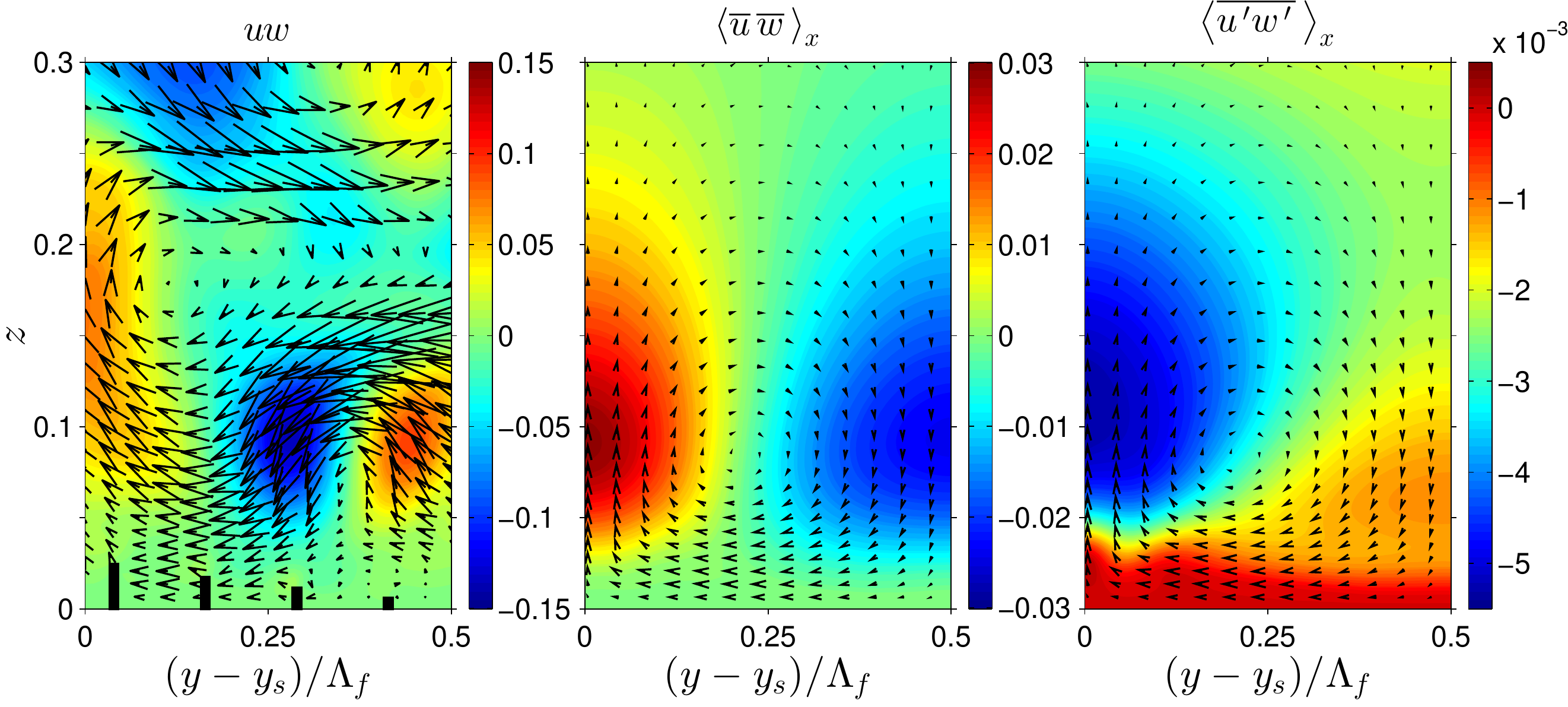}
\end{minipage}
\caption{Advective transport in a plane perpendicular to the streamwise direction for the herringbone texture with $\ngroove = 4, \ang = 165\degree$. Contours represent the instantaneous advection $uw$ (\textbf{left}), the averaged mean advection (\textbf{center}) or the averaged turbulent advection (\textbf{right}). Vectors depict the instantaneous (\textbf{left}) or averaged (\textbf{center}, \textbf{right}) in-plane flow velocity.}
\label{fig:uw_decomposition}
\end{figure*}

The variation of the mean and turbulent drag contributions with feather width very much resembles that of the mean and turbulent secondary flow strengths (compare \autoref{fig:FIK_herringbone} and \autoref{fig:secondary_flow_strength_vs_ngroove}). It thus appears that advection and secondary flow go hand in hand. The same can be observed from the instantaneous flow field in \autoref{fig:uw_decomposition}, which shows both contours of advection and vectors of secondary flow. 
Regions with strong secondary flow (in particular strong $w$) are also regions with strong advection.
Both secondary flow and the advective flux are split into mean and turbulent contributions in view of $\secflow^2 = \secflowmean^2 + \secflowturb^2$ and $\tav{uw} = \tav{u}\,\tav{w} + \tav{u'w'}$, respectively.
In summary, the results clearly indicate an intimate connection between increased advective transport and stronger secondary flow.


Given the close correspondence between secondary flow and advection, the change of the secondary flow strength with feather width as clarified in \autoref{sec:secondary_flow_strength} also explains the trend of the advective drag contributions. In particular, the increased advective drag for smaller feather widths is due to the higher spanwise density of the converging or diverging riblets that generate the secondary flows. Also, reduction of the mean advective drag for $\ngroove = 1$ is ascribed to a weaker mean secondary flow due to spanwise confinement.
There is, however, one major difference between \autoref{fig:FIK_herringbone} and \autoref{fig:secondary_flow_strength_vs_ngroove}, namely the effect of shifting of the top-wall texture. For 16 and 32 grooves, the mean secondary flow for the shifted textures is clearly stronger than for the not-shifted textures. In contrast, the drag due to mean advection is about the same, independent of the shift. This is explained by the fact that a stronger secondary flow near the channel centerline does not contribute much to drag because of the factor $\lb 1/2 - z \rb$ in \autoref{eq:almost_extended_FIK_identity}. This demonstrates that mean-secondary-flow strength is a good indicator for the contribution of mean advection to drag, provided that the mean secondary flow near the centerline is weak.

\begin{figure*}[t!]
\begin{minipage}[t]{1\linewidth}
\centering
\includegraphics[width=\textwidth]{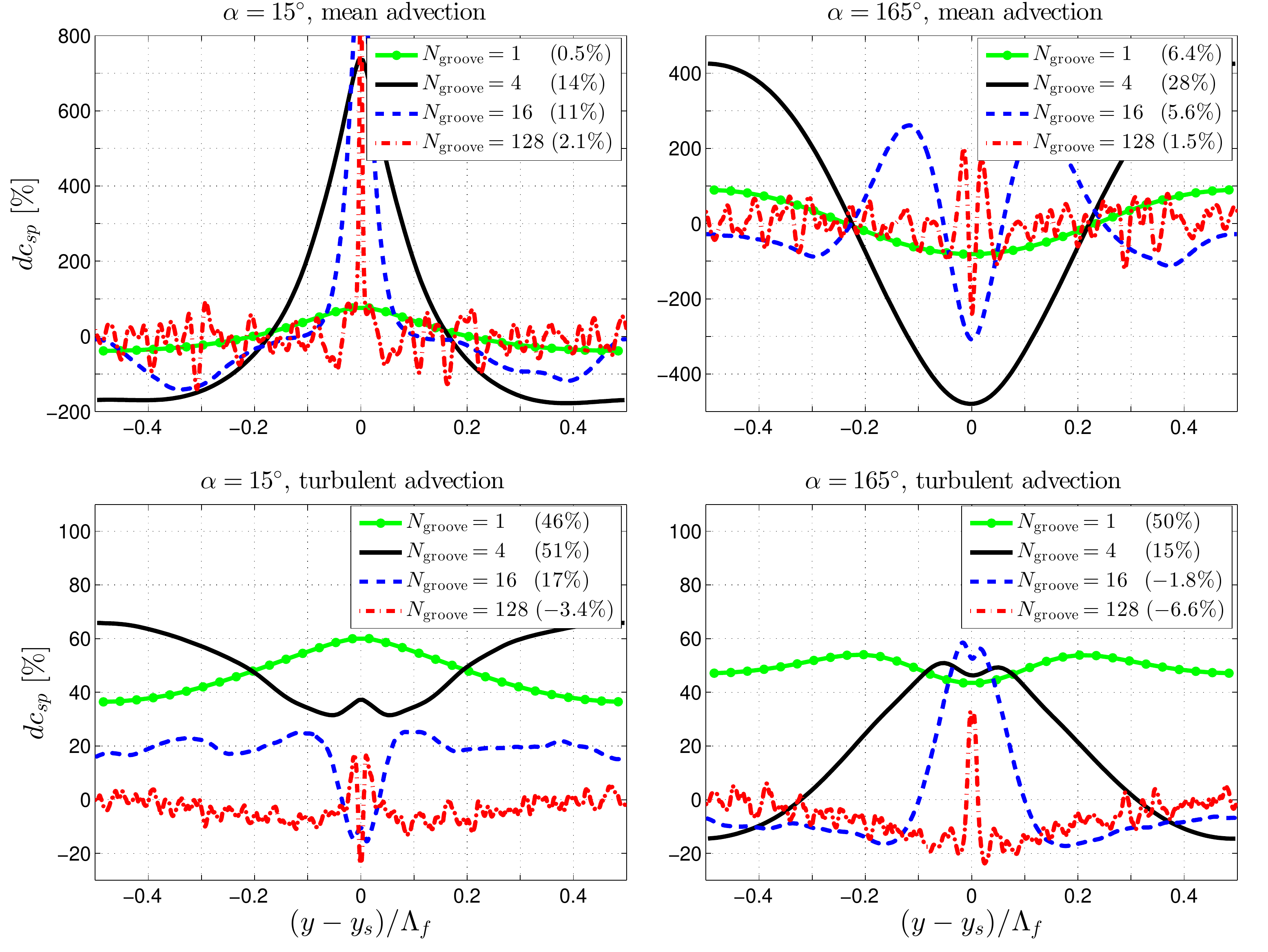}
\end{minipage}
\caption{Spanwise decomposition of advective contributions to the total drag change (\autoref{eq:DC_spanwise_decomposition}).
The total advection term is split into a mean (\textbf{top}) and turbulent (\textbf{bottom}) part. Results for herringbone textures with $\ang = 15\degree$ (\textbf{left}) and $\ang = 165\degree$ (\textbf{right}) are shown.
The number in the legend quantifies $dc$, the spanwise average of each curve (\autoref{eq:DC_spanwise_decomposition}), which was shown already in \autoref{fig:FIK_herringbone}.}
\label{fig:FIK_herringbone_vs_y}
\end{figure*}

\textbf{Spanwise decomposition of drag change}
%
%
%
\autoref{fig:FIK_herringbone} shows that drag reductions are possible in the limit of large $\ngroove$. Like for the conventional riblet texture, these reductions originate from weakened turbulent advective transport. To reveal the origin of this weakening, the spanwise dependence of the advective FIK-terms is investigated with a spanwise decomposition. Let $\avxy{\tav{f}}$ be one term in \autoref{eq:streamwise_momentum_balance_vs_z} and $d^{(l)}$ the corresponding drag contribution. The calculation of this FIK-term can be rewritten as:
\begin{equation}
\begin{aligned}
d^{(l)} & = \intn{0}{1}{ g \avxy{\tav{f}} }{z} \\
		& = \avy{ \intn{0}{1}{ g \avx{\tav{f}} }{z} } 
			\equiv \avy{ \dsp^{(l)} }.
\end{aligned}
\label{eq:FIK_spanwise_decomposition}
\end{equation}
The function $g = g(z)$ results from conversion of a triple to a single integral, and normalization. The function $\dsp^{(l)} = \dsp^{(l)}(y)$ represents the spanwise decomposition of the FIK-term, as indicated by the subscript $sp$. The drag change contribution $dc^{(l)}$ can be decomposed in a similar way:
\begin{equation}
\begin{aligned}
dc^{(l)}	& = \frac{d^{(l)} - \fsmooth^{(l)} }{\ftotsmooth} \\
			& = \avy{ \frac{ \dsp^{(l)} - \fsmooth^{(l)} }{\ftotsmooth} } 
			 	\equiv \avy{ \dcsp^{(l)} },
\end{aligned}
\label{eq:DC_spanwise_decomposition}
\end{equation}
where $\dcsp^{(l)}$ quantifies how the drag change depends on the spanwise coordinate. Note that $\dsp^{(l)}$ and $\dcsp^{(l)}$ have a spanwise dependence, whereas $\fsmooth^{(l)}$ and $\ftotsmooth$ have not.
The spanwise decomposition of the mean and turbulent advective terms is shown in \autoref{fig:FIK_herringbone_vs_y} for $\ang = 15\degree$ and $165\degree$, and four feather widths. Small asymmetries with respect to $y = y_s$ are attributed to slow convergence of weak secondary flows. The small oscillations in the curves for $\ngroove = 128$ probably appear for the same reason.

The effect of convergent/divergent riblets can be isolated from that of parallel riblets in yaw for the textures with $\ngroove = 128$. The parallel riblets dominate the flow in a region sufficiently far away from the shaft. In that region, the mean advection term fluctuates around zero. The turbulent term is zero at the feather edges. Riblets have no height there and the smooth-wall result is recovered. Away from the edges, the term decreases more or less linearly, which is attributed to riblet-height increase and the associated suppression of turbulent transport. It approaches approximately $-10$\%, which belongs to full-height parallel riblets with yaw angle $\ang = 15\degree$ or $165 \degree$. These favorable trends of mean and turbulent advective drag stop near the shaft because of the strong secondary flow there.


The converging/diverging riblets dominate the flow in a region around the shaft. That is especially evident from the drag change due to mean advection, which shows a clear signature of the mean flow described before, in particular the strong secondary-flow vortices near the shaft.
For $\ang = 165\degree$ and any $\ngroove$, the updraft of fluid around the shaft results in a local drag reduction, as is apparent from the dip in $\dcsp$ at $y = \ys$. The peak next to this dip is associated with that part of the vortex that transports momentum towards the wall. For $\ngroove = 16$, a second dip appears, which is ascribed to the tertiary flow shown in \autoref{fig:mean_secondary_flow_herringbone}. The central dip is narrower for larger $\ngroove$, because the secondary-flow vortex near the shaft is smaller compared to the total feather width. A similar (but opposite) description applies to textures with $\ang = 15\degree$.

Local turbulent drag reduction can be obtained by diverging riblets. For $\ang = 15\degree$ and $\ngroove \geq 16$, the strong wall-directed mean flow is accompanied by reduced turbulent transport. The reverse is true for $\ang = 165\degree$, namely that the strong wall-leaving mean flow is accompanied by increased wall-directed turbulent transport, as can also be seen from \autoref{fig:uw_decomposition}.
These observations fully agree with the findings for uniform blowing or suction. In Direct Numerical Simulations, it has been found that uniform blowing reduces mean advective drag and enhances turbulent drag, while uniform suction enhances mean advective drag and reduces turbulent drag \citep{kametani2011direct}.

The local contribution of advection to drag might be very different from its global (or volume-averaged) contribution. For instance, mean advection might seem much more important than turbulent advection in view of the scales of \autoref{fig:uw_decomposition} and \autoref{fig:FIK_herringbone_vs_y}. However, the volume-averaged turbulent term is often at least as important as the mean term. As a second example, the texture with $\ang = 165\degree$ and $\ngroove = 4$ exhibits locally a drag reduction as high as 400\% due to a strong wall-leaving flow. However, one should realize that the favorable updraft of low-momentum fluid is compensated by an adverse downdraft of high-momentum fluid. The influence of the whole vortex on the drag should be considered. In the case of $\ang = 165\degree$ and $\ngroove = 4$, the vortex covers one feather half. The corresponding spanwise-averaged mean advective drag is 28\%, which establishes again an unfavorable effect of the vortices near the shaft on the drag. Therefore, one should be careful to judge the performance of this (or any) texture based on a local drag determination. The mean or turbulent advective drag might be reduced locally. However, when the flow is dominated by strong advection (such as near the shaft), the volume-averaged drag generally increases due to an overall increase of both the mean and turbulent advective contributions.


In summary, the present study confirms two effects of the herringbone riblet texture on the turbulent drag. 
The first effect relates to the texture-generated secondary flows around the shaft. Although turbulent drag might be reduced locally due to a wall-directed mean flow, the overall trend is an increased turbulent drag because of the fluctuating secondary flows that are generated by the converging/diverging riblets near the shaft.
The second effect relates to the parallel riblets sufficiently far away from the shaft. Those riblets suppress turbulent transport, which results in turbulent drag reduction.




\section{Discussion} \label{sec:discussion}

\textbf{Detrimental effect of convergent/divergent riblets}
The results presented in this paper clarify the influence of the herringbone riblets on the drag.
The drag increase for $\wf/\lz \lesssim O(1)$ is attributed to an increase of advective transport (\autoref{fig:FIK_herringbone}), which in turn is associated with the secondary-flow vortex near the shaft (\autoref{fig:FIK_herringbone_vs_y}). Although the vortex might locally be responsible for a drag reduction, as a whole it is responsible for a drag increase.
That vortex originates from the herringbone riblets near the shaft (Figures \ref{fig:mean_secondary_flow_herringbone}, \ref{fig:secondary_flow_strength_vs_ngroove}, \ref{fig:mom_balance_vs_y}). That strongly suggests that these converging/diverging riblets are detrimental to DR. 
That is confirmed by the finding that $\wf \rightarrow \infty$ is most beneficial for DR (\autoref{fig:DC_vs_ngroove}). In that limit, the herringbone texture approaches the conventional parallel-riblet texture in yaw. Hence, the presence of convergent/divergent riblets in the texture seems unfavorable for DR.

\textbf{Research on spanwise-periodic spanwise forcing}
Research on the drag-reducing spanwise forcing further underpins the detrimental effect of the herringbone riblets on the drag. 
The spanwise traveling wave of spanwise body force is described by $f_y = A(z)\sin\{(2\pi/\wf)y - \omega t\}$. Here $f_y$ represents the spanwise body force, $A(z)$ the forcing amplitude (only nonzero in the vicinity of the wall), $\wf$ the spanwise wavelength of the forcing, and $\omega$ the angular frequency of the forcing.
The herringbone texture is similar to this forcing with respect to its spanwise periodicity, though the texture is static ($\omega = 0$).
Research shows that the best-performing spanwise traveling wave (in terms of DR or net energy saving) is the one with infinite wavelength, i.e. the spanwise wall oscillation \citep{du2002drag, quadrio2015turbulent}. In other words, $\wf \rightarrow \infty$ is most beneficial for DR, like for the herringbone texture. 
Hence, the addition of spanwise periodicity seems detrimental to the drag-reducing performance of both the spanwise forcing and the riblet texture in yaw.

\textbf{Alternative passive riblet textures}
As the herringbone texture with a spanwise variation of the forcing appears to be unprofitable for DR, a streamwise variation of the forcing might be more advantageous. Indeed, streamwise traveling waves of spanwise forcing are superior to spanwise traveling waves of spanwise forcing, presumably because of the unaltered wall-normal gradient of the wall-normal velocity component at the wall \citep{quadrio2015turbulent}. 
Both experimentally and numerically, DR has been obtained with a stationary streamwise variation of \textit{spanwise} forcing \citep{viotti2009, gruneberger2012wavelike}. DR by streamwise variation of \textit{wall-normal} forcing (e.g. suction and blowing) has been reported as well, although the net energy saving has been small \citep{quadrio2007effect, mamori2014drag}.
\change{
The drag increasing/reducing trends in the current study are similar to what has been found for uniform suction/blowing (Figures \ref{fig:instantaneous_vortical_structures}, \ref{fig:FIK_herringbone_vs_y}), but a net drag-reducing effect could not be confirmed. It is difficult (if not impossible) to eliminate the unfavorable effects of downdrafts, as mass conservation dictates that updrafts need to be compensated by downdrafts. Also, textures with converging/diverging riblets are not fully comparable to uniform blowing/suction. For instance, uniform blowing originates from a nonzero mass flux through the wall, in contrast to the texture-generated updrafts.}
Still, textures with streamwise variation of the forcing seem more promising than the herringbone texture.

\textbf{Comparison with experiments}
The conclusion that the herringbone texture seems detrimental to turbulent drag reduction apparently contrasts with the experimental study of \citet{chen2014flow}. Although it has not been the aim of the present study to reproduce their experiments numerically, a comparison might still be illuminating.
\citet{chen2014flow} obtained drag reduction in forward flow for $\wf/D \lesssim 1$ ($D$ being the pipe diameter), with a maximum of 20\%. In contrast, the present study only achieved DR for $\wf/\lz > 10$, with a maximum of 2\% in backward flow.
These contrasting results might be ascribed to differences in riblet texture and Reynolds number.

First, the textures were different, particularly the riblet shape and angle. The feather width might have been different too, but it is unclear what value for $\ngroove$ was used in the experiments.
The experimental texture consisted of sawtooth riblets at an angle of $30\degree$ with the flow direction. The numerical texture was composed of blades at an angle of $15\degree$. Although blades seem to be more sensitive to yaw (see \autoref{fig:drag_parallel_riblets}), they were studied at a smaller yaw angle. Therefore, there is currently no clear indication that the use of blade riblets in the numerical study contributed to a lesser drag-reducing performance of the herringbone texture.


Second, the Reynolds numbers differed significantly. The 20\% DR was obtained at $\rebulk \approx 2.6\e{5}$ (based on pipe diameter and bulk velocity), while the numerical study was performed at $\rebulk = 5500$. For DR techniques that rely on near-wall flow manipulation, the attainable DR is approximately independent of the Reynolds number (although not fully \citep{iwamoto2002reynolds, iwamoto2005friction, spalart2011drag, gatti2013performance}) when near-wall scaling is applied. However, herringbone riblets (in contrast to conventional riblets) cannot be regarded as viscous-region modifiers, because they generate a secondary flow throughout a large part of the channel (\autoref{fig:mean_secondary_flow_herringbone}). Therefore, viscous scaling is probably inappropriate in this case. Because of the unknown scaling and the Reynolds number difference, the texture parameters (e.g. $\wf$) and the drag reduction results of the experiments and the simulations cannot be directly compared.

\section{Conclusions and Outlook} \label{sec:conclusions}

The drag-reducing performance of a herringbone riblet texture was studied with Direct Numerical Simulations (DNSs) of turbulent flow in a channel with height $\lz$. The FIK-identity for drag decomposition was extended to textured walls and was used to study the underlying drag change mechanisms.
For validation, simulations with smooth walls, parallel blade riblets and parallel blade riblets in yaw were performed, which showed good agreement with literature. The parallel-riblet simulations exhibited an expected but small low-Reynolds-number effect. A maximum drag reduction (DR) of 9.3\% was obtained, close to the 9.9\% that has been found experimentally \citep{bechert1997}.


The herringbone texture can both increase or reduce the drag, depending on the spanwise texture wavelength $\wf$. For $\wf/\lz \lesssim O(1)$ (i.e. narrow feathers), the drag increases with a maximum of 73\% for $\wf/\lz = 0.4$. This increase is ascribed to the convergent/divergent riblets. They generate a fluctuating secondary flow, which on average consists of two counter-rotating vortices centered above the regions of riblet convergence/divergence. The strong secondary flow increases both mean and turbulent advective transport, which in turn results in the significant drag increase.


A slight drag reduction of 2\% was found for $\wf/\lz \gtrsim O(10)$ (wide feathers). Due to the large feather width, the secondary flow generated by the converging/diverging riblets now influences only a relatively small part of the whole texture. Its drag-increasing contribution is therefore small. The largest part of the texture behaves similarly to a conventional parallel-riblet texture in yaw. Specifically, suppression of turbulent advective transport is responsible for the small DR that was obtained.

As was found by other researchers for spanwise traveling waves of spanwise forcing, the current study confirms that $\wf \rightarrow \infty$ is most beneficial for DR. In that limit, the texture approaches the conventional parallel-riblet texture in yaw. Therefore, the presence of convergent/divergent riblets in the texture seems detrimental to turbulent drag reduction, which apparently contrasts with the experiments of \citet{chen2014flow}. However, differences in Reynolds number and texture parameters (riblet shape, feather width, angle between riblets and flow direction) hindered a one-to-one comparison between the present simulations and the experiments.

More elaborate experiments and simulations are required to further investigate the drag-reducing potential of the herringbone texture. First, the maximum DR of 20\% should be reproduced and the optimum texture parameters ($\wf$, $s$, $h/s$, $\ang$) should be determined. In numerical simulations, the more realistic sawtooth riblet geometry should be implemented. The Reynolds-number influence and the parameter scaling also need further attention. As the herringbone riblets generate a secondary flow throughout a large part of the channel, the scaling of drag reduction with the texture and flow parameters is presumably non-trivial.

Whether feather riblets have an aerodynamic function remains an open question. \citet{chen2014flow} claimed that feather riblets greatly impact flight performance because of drag reduction. The DR was ascribed to suppression of turbulent momentum transport, but evidence has been inconclusive so far. One should realize that wings of birds are not flat and operate at a relatively low Reynolds number, so flow-separation delay seems a more plausible aerodynamic function of feather riblets. Indeed, several studies confirm that roughness on a bird wing contributes to separation control \citep{bushnell1991nature, lilley1998study, bokhorst2015feather}. Furthermore, separation delay has been obtained with vortex generators that resemble the herringbone texture \citep{lin2002review}. So, future studies might investigate the potential of the herringbone texture for flow-separation control.

\section*{Acknowledgements}
The research leading to these results has received funding from the European Union Seventh Framework Programme in the SEAFRONT project under grant agreement nr. 614034.
The simulations were performed on the Dutch national supercomputer Cartesius at SURFsara, Amsterdam, The Netherlands. This work was sponsored by NWO Physical Sciences for the use of supercomputer facilities.
H.O.G.B. would like to thank his colleague Pedro Costa for his supercomputational support.

\bibliographystyle{abbrvnat}



\appendix


\change{

\section{Time Advancement at Fixed Bulk Velocity} \label{sec:numerical_code}

For simplicity, \autoref{eq:Navier_Stokes_equations} is rewritten as:
\begin{align}
\pd{u_i}{t} & = -\pd{p}{x_i} + r_i + \gammaf \fdriving \delta_{i1},
\end{align}
where $r_i$ contains the advection, diffusion and IBM terms. Let the integer $n$ denote the time steps. The used Runge-Kutta scheme (RK3) employs three sub-steps, which are numbered by the integer $s$. RK3 introduces intermediate velocities $\uintmed{i}{s}$, where $\uintmed{i}{0} = u_i^n$ and $\uintmed{i}{3} = u_i^{n+1}$. Similarly, intermediate pressures $\pintmed{s}$ and driving forces $\fdrivingintmed{s}$ are introduced. Because of the Crank-Nicolson scheme for pressure, $\pintmed{s} = \pintmed{s-1} + \pcor{s}$ with correction pressure $\pcor{s}$. The time advancement is illustrated here for an arbitrary sub-step $s$:
\begin{align}
\uintmed{i}{s} + \alphas{s} \dt \pd{\pcor{s}}{x_i} & = 
\uintmed{i}{s-1} - \alphas{s} \dt \pd{\pintmed{s-1}}{x_i} + 
\dt \lb \gammas{s} \rintmed{i}{s-1} + \zetas{s-1} \rintmed{i}{s-2} \rb + \nonumber \\
& \hspace{2cm} \alphas{s} \dt \gammaf \fdrivingintmed{s} \delta_{i1} \nonumber \\
& \equiv \upredone{i}{s} + \alphas{s} \dt \gammaf \fdrivingintmed{s} \delta_{i1} \label{eq:numerical_scheme} \\
& \equiv \upredtwo{i}{s}, \nonumber
\end{align}
where $\upredone{i}{s}$ is the first and $\upredtwo{i}{s}$ the second prediction velocity. Note that the asterisk here is not used to denote dimensional quantities. The parameters $\alphas{s}$, $\gammas{s}$ and $\zetas{s-1}$ are RK3 parameters (see e.g. \cite{wesseling2001principles}). To obtain the driving force, the equation for the streamwise velocity is volume-averaged, which yields:

%
\begin{equation}
\begin{aligned}
\vav{\uintmed{}{s}} + \alphas{s} \dt \vav{\pd{\pcor{s}}{x}} = 
 \vav{\upredone{}{s}} + \alphas{s} \dt \vav{\gammaf} \fdrivingintmed{s}.
\end{aligned}
\end{equation}
The first term represents the intermediate bulk velocity in sub-step $s$, which is set equal to one to obtain the constant bulk flow. The second term disappears because of periodic boundary conditions. The first prediction velocity is known, so the third term can be computed. The resulting equation can be solved for $\fdrivingintmed{s}$, which yields:
\begin{equation}
\begin{aligned}
\fdrivingintmed{s} = \frac{1 - \vav{\upredone{}{s}}}{ \alphas{s} \dt \vav{\gammaf} }.
\end{aligned}
\end{equation}
Next, $\upredtwo{i}{s}$ is computed. With use of the continuity equation, the divergence of \autoref{eq:numerical_scheme} yields a Poisson equation for the correction pressure:
\begin{equation}
\begin{aligned}
\pds{\pcor{s}}{x_j} = \frac{1}{\alphas{s} \dt} \pd{\upredtwo{j}{s}}{x_j}.
\end{aligned}
\end{equation}
When this is solved, the updated velocity and pressure are computed:
\begin{equation}
\begin{aligned}
\uintmed{i}{s} = \upredtwo{i}{s} - \alphas{s} \dt \pd{\pcor{s}}{x_i}, \hspace{1cm}
\pintmed{s} = \pintmed{s-1} + \pcor{s}.
\end{aligned}
\label{eq:correction_step}
\end{equation}
%
This procedure guarantees that the bulk velocity equals one in each RK3 sub-step, which results in three values of the driving force. The total forcing (i.e. one per time step) is calculated as follows:
\begin{equation}
\begin{aligned}
\fdriving = \sum_{s = 1}^{3} \alphas{s} \fdrivingintmed{s}.
\end{aligned}
\label{eq:sum_RK3_forcings}
\end{equation}
%

}

\section{Immersed Boundary Method for Blade Riblet Textures} \label{sec:IBM}

The IBM adjusts diffusive and advective fluxes around blades with help of indicator functions, as is described below for parallel riblets without and in yaw. 
\change{ The third subsection explains how the $\fibm_i$ term in \autoref{eq:Navier_Stokes_equations} is computed. } 
The final subsection describes how well the IBM approximates the no-slip and no-penetration conditions at the texture surface.


\subsection{Parallel Riblets without Yaw}

\begin{figure*}[t!]
\begin{minipage}[t]{1\linewidth}
\centering
\includegraphics[width=\textwidth]{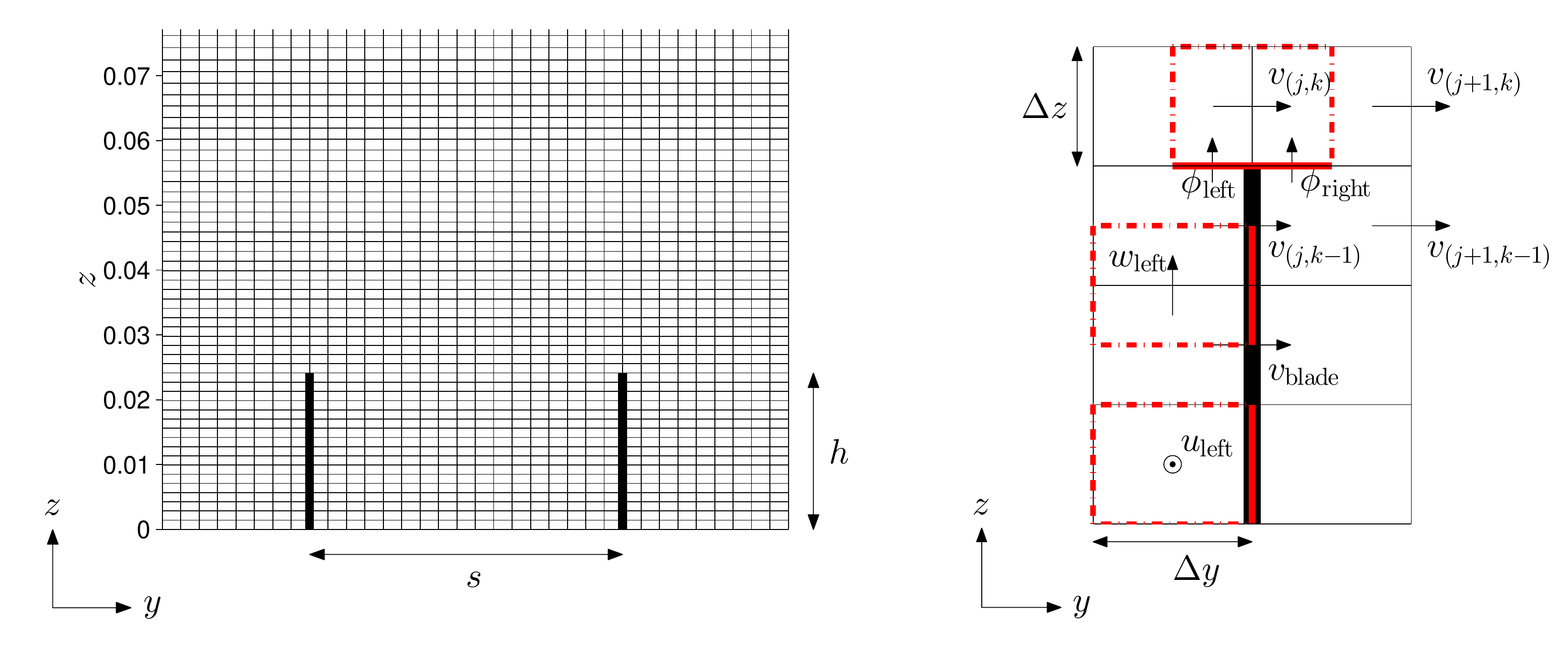}
\end{minipage}
\caption{Numerical grid and IBM used for parallel riblets without yaw.
\textbf{Left:} Part of the numerical grid used for $\splus = 17$ simulations at $\rebulk = 5500$, showing two unit cells in the spanwise direction. Both for blade spacing and blade height, 17 grid cells were used.
\textbf{Right:} Small part of the numerical grid (8 grid cells) around the blade tip. The text explains the IBM with use of this subfigure. The three red boxes represent staggered grid cells that belong to the velocity components shown in their centers. Each of these cells has one face indicated with a thick and solid line. At that cell face, the IBM adjusts the advective and diffusive fluxes.}
\label{fig:grid_blade_IBM}
\end{figure*}

\autoref{fig:grid_blade_IBM} shows the numerical grid used for parallel riblets without yaw. As a staggered grid is used, the velocity vectors are located at the faces of the grid cells. A few of these vectors are shown in the right figure. The blade coincides with a few grid points of the spanwise velocity $v$. 
The right figure helps to explain the IBM. The red boxes in that figure indicate three staggered grid cells that belong to the velocity components $u_\mathrm{left}$, $v_{(j,k)}$ and $w_\mathrm{left}$. The IBM implements an adjustment of the advective and diffusive fluxes at the cell faces marked with a thick solid line, as is explained below.


The streamwise velocity $u$ was only adjusted in grid cells next to the blades (both on the left and right side). For the grid cell of $u_\mathrm{left}$, the two spanwise fluxes through its thick solid face were changed: the advective flux $uv = 0$ and the diffusive flux $\partial u / \partial y = -2 u_\mathrm{left}/\dy$.


The spanwise velocity $v$ was changed for two reasons, namely to enforce no-penetration at the blades and to adjust the fluxes above the blade tip.
To enforce no-penetration, the prediction velocity was set to zero at the grid points that coincide with a blade. Let $v_\mathrm{blade}$ represent a spanwise velocity component that coincides with a blade (see \autoref{fig:grid_blade_IBM}) and let $^*$ represent the first prediction velocity (as in \autoref{eq:numerical_scheme}), then $v_\mathrm{blade}^* = 0$. The actual velocity $v_\mathrm{blade}$ follows from the correction step (\autoref{eq:correction_step}) and is very close to zero, although not exactly zero (details follow in \autoref{sec:BC}).


The second change to $v$ comprises the adjustment of advective and diffusive fluxes in the grid cell just above the blade tip (shown in red in \autoref{fig:grid_blade_IBM}). The vertical fluxes of $v$ at the bottom face of that cell were adjusted. The diffusive flux $\partial v / \partial z$ was split into two contributions, namely from the left and right side of the blade (indicated by $\phi_\mathrm{left}$ and $\phi_\mathrm{right}$ in the figure). It accounts for the fact that the thin blade does not inhibit vertical transport. Specifically, $\partial v / \partial z = 0.5 \phi_\mathrm{left} + 0.5 \phi_\mathrm{right}$. 
The flux $\phi_\mathrm{right}$ was computed by linear interpolation of the four velocity components labeled with indices in the figure: $\phi_\mathrm{right} = (v_t - v_b)/\dz$ with $v_t = 0.75 v_{(j,k)} + 0.25 v_{(j+1,k)}$ and $v_b = 0.75 v_{(j,k-1)} + 0.25 v_{(j+1,k-1)}$.
The flux $\phi_\mathrm{left}$ was computed in a similar way.
As advection near the blade tips is likely less important than diffusion, the advective flux was not split but simply set to zero (i.e. $vw = 0$ at the thick solid face).

The wall-normal velocity $w$ was only adjusted in grid cells next to the blades, similarly to what was done for $u$. For the grid cell of $w_\mathrm{left}$, two spanwise fluxes at the thick solid face were changed: the advective flux $vw = 0$ and the diffusive flux $\partial w / \partial y = -2 w_\mathrm{left}/\dy$. These adjustments were also applied for the $w$-cell next to the blade tip, so this diffusive flux was not split into two contributions (although the blade covers only half of the cell face). To justify this choice, the simulation with $\splus = 24$ at $\rebulk = 5500$ was repeated. The diffusive flux near the blade tip was separated into two contributions, namely from above and below the blade tip. No significant difference in drag was found.

\subsection{Parallel Riblets in Yaw}

\begin{figure*}[t!]
\begin{minipage}[t]{1\linewidth}
\centering
\includegraphics[width=\textwidth]{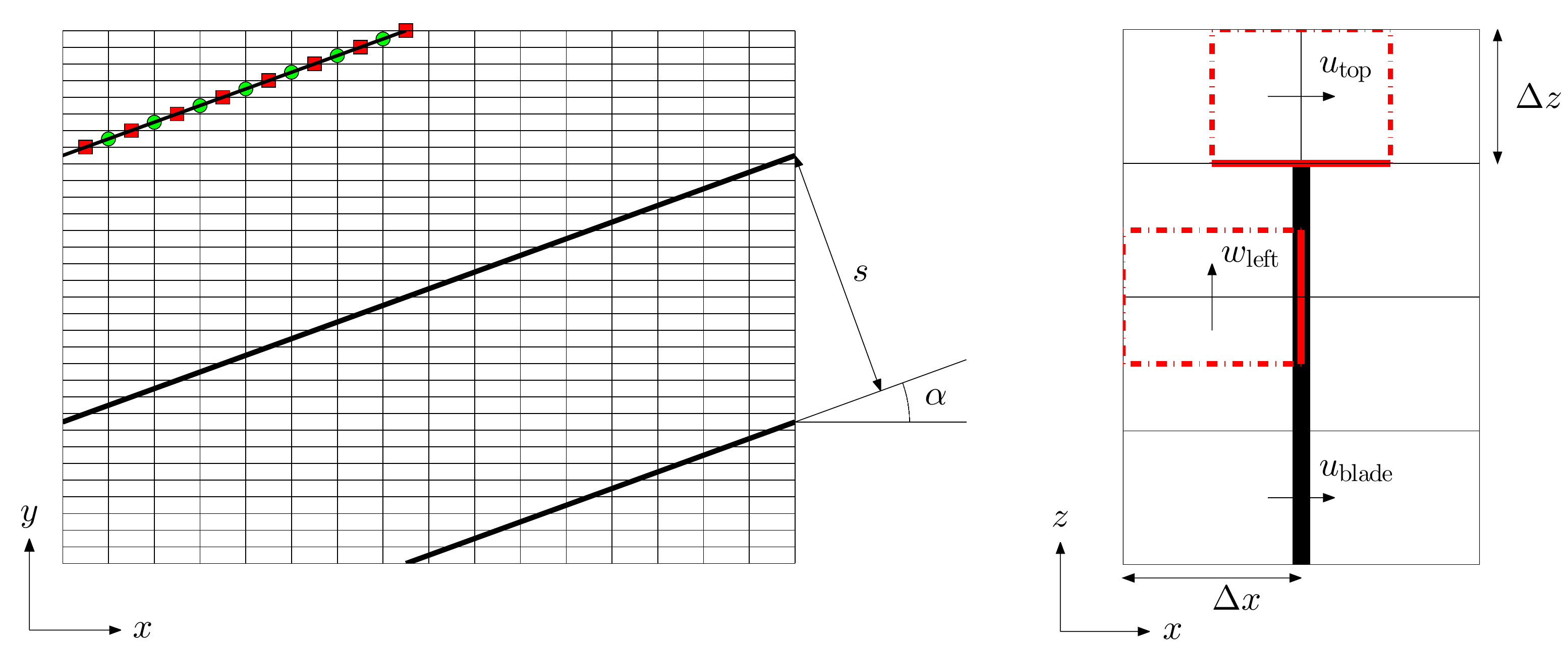}
\end{minipage}
\caption{Numerical grid and IBM used for parallel riblets in yaw.
\textbf{Left:} Top view of the texture with $\alpha = 20\degree$, showing one unit cell in the streamwise and two in the spanwise direction. The numerical grid consists of $\ncg = 16$ grid cells per groove.
The markers on one blade indicate that the blades intersect the grid cells at the locations of the staggered velocity vectors $u$ (green circle) and $v$ (red square).
\textbf{Right:} Small part of the numerical grid (8 grid cells) around the blade tip. The text explains the IBM with use of this subfigure, see also \autoref{fig:grid_blade_IBM}.}
\label{fig:yaw_texture}
\end{figure*}

To simulate the turbulent flow over blades in yaw, the riblet texture was rotated with respect to the grid, see \autoref{fig:yaw_texture} (left). That required a different Immersed Boundary Method (IBM), as the surfaces were not anymore aligned with the Cartesian directions. As the left figure shows, the grid was generated in such a way that the obstacle surface still coincides with part of the staggered velocity vectors. The main disadvantage of this approach is that the streamwise and spanwise grid spacings cannot be chosen independently: they depend on $\ang$. However, the major advantage is a relatively simple IBM. 
Due to the specific alignment of grid and texture, the number of grid cells per groove $\ncg$ is equal for the streamwise and spanwise directions. The grid shown in \autoref{fig:yaw_texture} has $\ncg = 16$.

The IBM is explained with use of \autoref{fig:yaw_texture} (right). Staggered grid cells for $u_\mathrm{top}$ and $w_\mathrm{left}$ are shown as red boxes. The IBM implements an adjustment of the advective and diffusive fluxes at the cell faces marked with a thick solid line, as is explained below.

The streamwise and spanwise velocities were adjusted in a similar way, so only the change to $u$ is described here.
The changes to $u$ were almost identical to that for $v$ in the previous subsection. 
To enforce no-penetration, the streamwise prediction velocity was set to zero (i.e. $u_\mathrm{blade}^* = 0$) at the grid points that coincide with a blade.
At the bottom cell face of the grid cell for $u_\mathrm{top}$, both the advective and diffusive fluxes were modified. 
The advective flux was set to zero: $uw = 0$. 
The diffusive flux was not split into two contributions: $\partial u / \partial z = 2 u_\mathrm{top} / \dz$.


For the wall-normal velocity, two indicator functions were used, namely one for streamwise and another one for spanwise transport of wall-normal momentum. As the treatment of streamwise and spanwise transport of $w$ is analogous, only the former is described here. The wall-normal prediction velocity was not changed directly, as $w$ never coincides with blades. Therefore, only the fluxes in $w$-cells next to the blades were modified. Specifically, the fluxes at the right cell face of the grid cell for $w_\mathrm{left}$ were adjusted as follows: $uw = 0$ and $\partial w / \partial x = -2 w_\mathrm{left} / \dx$.

\begin{figure}[t!]
\centering
\includegraphics[width=\textwidth]{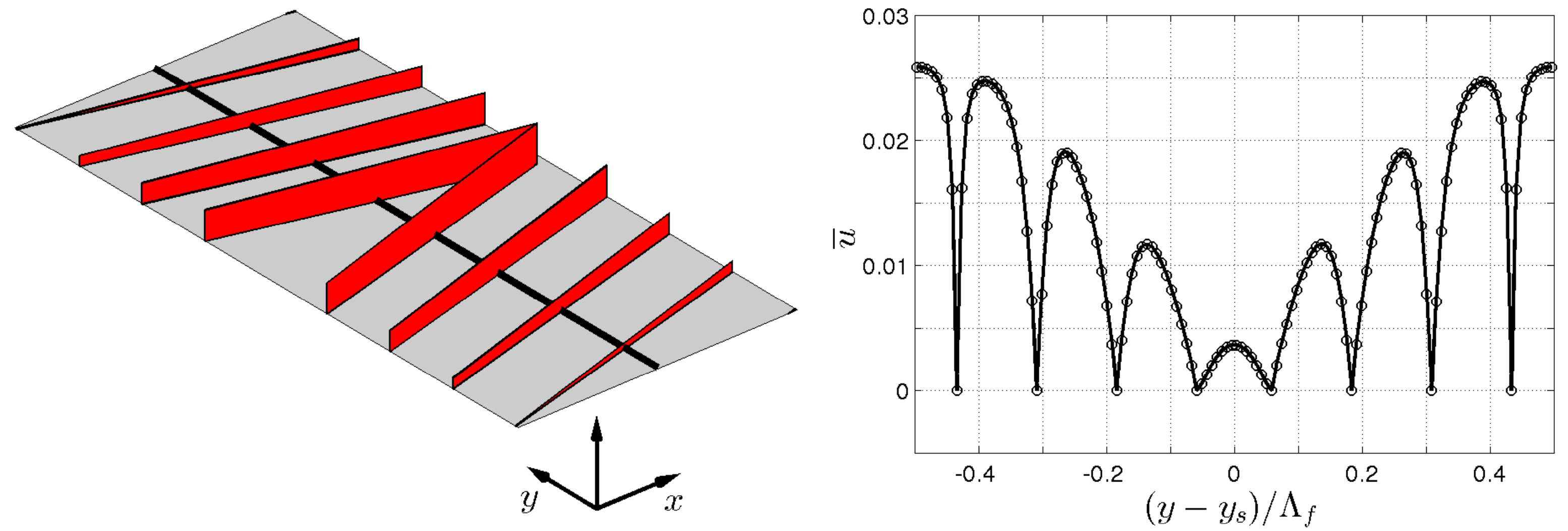}
\caption{
Spanwise profile of streamwise velocity. \textbf{Left:} Bird's-eye view of one unit cell of a herringbone texture with $\ngroove = 4$, $\alpha = 165\degree$. The thick line parallel to the $y$-axis cuts through the center of the first grid cell above the bottom wall. 
\textbf{Right:} Time-averaged streamwise velocity as function of spanwise distance along the thick line in the left subfigure.
}
\label{fig:streamwise_velocity_profile}
\end{figure}


\subsection{Computation of $\fibm_i$}

Although the IBM forcing $\fibm_i$ appears as a separate term in \autoref{eq:Navier_Stokes_equations} and following equations, it is not implemented as an explicit relation. The two previous subsections have shown instead that the numerical code for smooth walls was adjusted to obtain the zero velocity at the blades. However, the IBM term was needed as a separate term for the postprocessing. Its computation is illustrated with reference to \autoref{sec:numerical_code}. The first prediction velocity $\upredone{i}{s}$ that appears there contains all adjustments that are needed to obtain the zero velocity at the blades. Next, all these adjustments are left out to compute the prediction velocity $\upredoneNOIBM{i}{s}$ in absence of the IBM forcing. The intermediate IBM forcing $\fIBMintmed{i}{s}$ then follows from $\upredone{i}{s} = \upredoneNOIBM{i}{s} + \alphas{s} \dt \fIBMintmed{i}{s}$. Finally, the total IBM forcing $\fibm_i$ is obtained from summing $\alphas{s} \fIBMintmed{i}{s}$, similar to \autoref{eq:sum_RK3_forcings}.



\subsection{Boundary Conditions at the Texture Surface} \label{sec:BC}

The no-slip and no-penetration conditions at the texture surface need to be sufficiently satisfied for the IBM to be accurate. Let a penetration velocity denote the absolute value of a texture-collocated velocity component. As $u_\mathrm{blade}$ and $v_\mathrm{blade}$ denote two velocity components that coincide with the texture, then the corresponding penetration velocities are $\upen = \abs{u_\mathrm{blade}}$ and $\vpen = \abs{v_\mathrm{blade}}$. Note that $\wpen$ does not exist, as $w$ does not coincide with the textures studied in the present work. To check whether the no-penetration condition was sufficiently satisfied, each tenth simulation time-step three penetration-related velocities were computed, namely $\avs{\upen}$, $\avs{\vpen}$ and $\max{(\upen,\vpen)}$, where $\avs{...}$ denotes an average. The resulting time series were time averaged, yielding $\tav{\avs{\upen}}$, $\tav{\avs{\vpen}}$ and $\tav{\max{(\upen,\vpen)}}$. The magnitude of these penetration velocities (normalized with the bulk velocity) is similar for all the simulations, namely a mean on the order of $10^{-6}$ to $10^{-5}$ and a maximum on the order of $10^{-4}$ to $10^{-3}$. This shows that the penetration velocities are very small, so the no-slip and no-penetration conditions are sufficiently satisfied.

\autoref{fig:streamwise_velocity_profile} presents a streamwise velocity profile as function of the spanwise distance for a herringbone riblet texture with 4 grooves per feather half (so $\ngroove = 4$). The left subfigure shows the texture together with a line parallel to the $y$-axis. The right subfigure shows the time-averaged streamwise velocity profile that was extracted along that line. Clearly, the velocity is zero at the 8 riblet locations. This illustrates that the boundary conditions at the riblet surfaces are satisfied.
 

\section{Simulation Parameters} \label{sec:simulation_parameters}

An overview of all simulations with the corresponding parameters can be found in \autoref{tab:simulation_parameters}. 

\renewcommand{\arraystretch}{1.2}
\begin{table*}[htbp]
  \centering
  \begin{minipage}[t]{1\linewidth}
  \caption{Parameters of all Direct Numerical Simulations presented in this paper. 
  The simulations are grouped based on the type of texture. The short name indicates which parameters have been varied for a certain texture. The addition (sp. res.) refers to a case with double spanwise resolution; (res.) indicates a double streamwise and spanwise resolution.
  The parallel riblet variant with $\ngroove = 0.5$ refers to the conventional parallel-riblet texture.
  The averaging time $T$ is normalized with $\delta/\utauw$ derived from smooth-wall flow at the same bulk Reynolds number.}
  \resizebox{\columnwidth}{!}{
    \begin{tabular}{ l l l l l l l l l l l l l l l l l l l }
    \hline

short name	 & $\splus$	 & $\ang$ [$\degree$]	 & $h/s$	 & $\ngroove$	 & $\lx$	 & $\ly$	 & $\nx$	 & $\ny$	 & $\nz$	 & $\dxplus$	 & $\dyplus$	 & $\dzwplus$	 & $\dzcplus$	 & $\ncg$	 & $\rebulk$	 & $T\utauw/\delta$	 & $\dtot \e{3}$	 & $\dc$ [\%] \\ 

\hline
\multicolumn{19}{c}{\textbf{Smooth Wall}} \\
\hline 

	Re5500	 & -	 & -	 & -	 & -	 & 5.8	 & 2.9	 & 512	 & 512	 & 320	 & 4.0	 & 2.0	 & 0.50	 & 1.7	 & -	 & 5500	 & 136	 & 8.10 $\pm$ 0.02	 & 0.0 $\pm$ 0.4 \\ 
	Re11000	 & -	 & -	 & -	 & -	 & 4.9	 & 2.5	 & 800	 & 800	 & 512	 & 4.0	 & 2.0	 & 0.50	 & 2.0	 & -	 & 11000	 & 66	 & 6.81 $\pm$ 0.02	 & 0.0 $\pm$ 0.5 \\ 
	Re22000	 & -	 & -	 & -	 & -	 & 5.4	 & 2.7	 & 1600	 & 1600	 & 1024	 & 3.9	 & 2.0	 & 0.49	 & 1.8	 & -	 & 22000	 & 14	 & 5.70 $\pm$ 0.02	 & 0.0 $\pm$ 0.6 \\ 
	
\hline
\multicolumn{19}{c}{\textbf{Parallel Riblets}} \\
\hline
	
	splus10	 & 9.9	 & 0	 & 0.5	 & -	 & 4.1	 & 2.7	 & 360	 & 960	 & 320	 & 4.0	 & 1.0	 & 0.50	 & 1.8	 & 10	 & 5500	 & 95	 & 7.54 $\pm$ 0.04	 & -6.9 $\pm$ 0.5 \\ 
	splus17	 & 16.9	 & 0	 & 0.5	 & -	 & 4.1	 & 2.9	 & 360	 & 1020	 & 320	 & 4.0	 & 1.0	 & 0.50	 & 1.8	 & 17	 & 5500	 & 120	 & 7.49 $\pm$ 0.05	 & -7.6 $\pm$ 0.7 \\ 
	splus24	 & 23.9	 & 0	 & 0.5	 & -	 & 4.1	 & 2.7	 & 360	 & 960	 & 320	 & 4.0	 & 1.0	 & 0.50	 & 1.9	 & 24	 & 5500	 & 113	 & 8.34 $\pm$ 0.04	 & 3.0 $\pm$ 0.6 \\ 
	splus24 (sp. res.)	 & 23.9	 & 0	 & 0.5	 & -	 & 4.1	 & 2.7	 & 360	 & 1920	 & 320	 & 4.0	 & 0.5	 & 0.50	 & 1.9	 & 48	 & 5500	 & 69	 & 8.38 $\pm$ 0.04	 & 3.4 $\pm$ 0.6 \\ 
	splus17 Re11000	 & 16.8	 & 0	 & 0.5	 & -	 & 4.3	 & 2.9	 & 704	 & 1904	 & 512	 & 4.0	 & 1.0	 & 0.50	 & 2.1	 & 17	 & 11000	 & 39	 & 6.17 $\pm$ 0.03	 & -9.3 $\pm$ 0.5 \\ 
	splus24 Re11000	 & 23.8	 & 0	 & 0.5	 & -	 & 4.2	 & 2.7	 & 680	 & 1728	 & 512	 & 4.0	 & 1.0	 & 0.50	 & 2.2	 & 24	 & 11000	 & 50	 & 6.78 $\pm$ 0.04	 & -0.5 $\pm$ 0.7 \\ 
	splus24 Re22000	 & 24.0	 & 0	 & 0.5	 & -	 & 4.1	 & 2.6	 & 1200	 & 3072	 & 1024	 & 4.0	 & 1.0	 & 0.50	 & 1.9	 & 24	 & 22000	 & 13	 & 5.61 $\pm$ 0.04	 & -1.6 $\pm$ 0.8 \\ 
	
\hline
\multicolumn{19}{c}{\textbf{Parallel Riblets in Yaw}} \\
\hline
	
	alpha10	 & 16.9	 & 10	 & 0.5	 & -	 & 4.4	 & 2.5	 & 384	 & 1200	 & 320	 & 4.1	 & 0.7	 & 0.50	 & 1.8	 & 24	 & 5500	 & 124	 & 7.77 $\pm$ 0.05	 & -4.0 $\pm$ 0.7 \\ 
	alpha15	 & 16.9	 & 15	 & 0.5	 & -	 & 4.5	 & 2.5	 & 384	 & 800	 & 320	 & 4.1	 & 1.1	 & 0.50	 & 1.8	 & 16	 & 5500	 & 124	 & 8.02 $\pm$ 0.03	 & -0.9 $\pm$ 0.4 \\ 
	alpha15 (res.)	 & 16.9	 & 15	 & 0.5	 & -	 & 4.5	 & 2.5	 & 768	 & 1600	 & 320	 & 2.0	 & 0.5	 & 0.50	 & 1.8	 & 32	 & 5500	 & 133	 & 8.01 $\pm$ 0.04	 & -1.2 $\pm$ 0.5 \\ 
	alpha20	 & 16.9	 & 20	 & 0.5	 & -	 & 4.5	 & 2.6	 & 512	 & 800	 & 320	 & 3.1	 & 1.1	 & 0.50	 & 1.8	 & 16	 & 5500	 & 124	 & 8.40 $\pm$ 0.04	 & 3.7 $\pm$ 0.6 \\ 




\hline
\multicolumn{19}{c}{\textbf{Herringbone Riblets: $\boldsymbol{\alpha = 15\degree}$}} \\
\hline
	
	ngroove1	 & 17.0	 & 15	 & 0.5	 & 1	 & 4.7	 & 3.2	 & 400	 & 1024	 & 320	 & 4.1	 & 1.1	 & 0.50	 & 1.8	 & 16	 & 5500	 & 77	 & 12.46 $\pm$ 0.06	 & 53.8 $\pm$ 0.8 \\ 
	ngroove4	 & 17.0	 & 15	 & 0.5	 & 4	 & 4.7	 & 3.2	 & 400	 & 1024	 & 320	 & 4.1	 & 1.1	 & 0.50	 & 1.8	 & 16	 & 5500	 & 82	 & 14.0 $\pm$ 0.1	 & 73.4 $\pm$ 1.4 \\ 
	ngroove4 (res.)	 & 17.0	 & 15	 & 0.5	 & 4	 & 4.7	 & 3.2	 & 800	 & 2048	 & 320	 & 2.1	 & 0.5	 & 0.50	 & 1.8	 & 32	 & 5500	 & 46	 & 14.24 $\pm$ 0.09	 & 75.7 $\pm$ 1.3 \\ 
	ngroove16	 & 17.0	 & 15	 & 0.5	 & 16	 & 4.7	 & 3.2	 & 400	 & 1024	 & 320	 & 4.1	 & 1.1	 & 0.50	 & 1.8	 & 16	 & 5500	 & 72	 & 10.86 $\pm$ 0.05	 & 34.1 $\pm$ 0.7 \\ 
	ngroove128	 & 17.0	 & 15	 & 0.5	 & 128	 & 4.7	 & 12.9	 & 400	 & 4096	 & 320	 & 4.1	 & 1.1	 & 0.50	 & 1.8	 & 16	 & 5500	 & 77	 & 8.29 $\pm$ 0.02	 & 2.3 $\pm$ 0.4 \\ 
	
\hline
\multicolumn{19}{c}{\textbf{Herringbone Riblets: $\boldsymbol{\alpha = 165\degree}$}} \\
\hline
	
	ngroove1	 & 17.0	 & 165	 & 0.5	 & 1	 & 4.7	 & 3.2	 & 400	 & 1024	 & 320	 & 4.1	 & 1.1	 & 0.50	 & 1.8	 & 16	 & 5500	 & 77	 & 13.1 $\pm$ 0.1	 & 61.2 $\pm$ 1.5 \\ 
	ngroove4	 & 17.0	 & 165	 & 0.5	 & 4	 & 4.7	 & 3.2	 & 400	 & 1024	 & 320	 & 4.1	 & 1.1	 & 0.50	 & 1.8	 & 16	 & 5500	 & 75	 & 11.98 $\pm$ 0.07	 & 47.9 $\pm$ 1.0 \\ 
	ngroove16	 & 17.0	 & 165	 & 0.5	 & 16	 & 4.7	 & 3.2	 & 400	 & 1024	 & 320	 & 4.1	 & 1.1	 & 0.50	 & 1.8	 & 16	 & 5500	 & 77	 & 8.71 $\pm$ 0.04	 & 7.5 $\pm$ 0.5 \\ 
	ngroove32	 & 17.0	 & 165	 & 0.5	 & 32	 & 4.7	 & 3.2	 & 400	 & 1024	 & 320	 & 4.1	 & 1.1	 & 0.50	 & 1.8	 & 16	 & 5500	 & 73	 & 8.27 $\pm$ 0.04	 & 2.1 $\pm$ 0.5 \\ 
	ngroove128	 & 17.0	 & 165	 & 0.5	 & 128	 & 4.7	 & 12.9	 & 400	 & 4096	 & 320	 & 4.1	 & 1.1	 & 0.50	 & 1.8	 & 16	 & 5500	 & 77	 & 7.94 $\pm$ 0.02	 & -2.0 $\pm$ 0.4 \\ 
	
\hline
\multicolumn{19}{c}{\textbf{Herringbone Riblets: $\boldsymbol{\alpha = 165\degree}$, shifted variant}} \\
\hline
	
	ngroove1	 & 17.0	 & 165	 & 0.5	 & 1	 & 4.7	 & 3.2	 & 400	 & 1024	 & 320	 & 4.1	 & 1.1	 & 0.50	 & 1.8	 & 16	 & 5500	 & 76	 & 13.0 $\pm$ 0.1	 & 60.3 $\pm$ 1.3 \\ 
	ngroove4	 & 17.0	 & 165	 & 0.5	 & 4	 & 4.7	 & 3.2	 & 400	 & 1024	 & 320	 & 4.1	 & 1.1	 & 0.50	 & 1.8	 & 16	 & 5500	 & 76	 & 11.9 $\pm$ 0.1	 & 47.2 $\pm$ 1.4 \\ 
	ngroove16	 & 17.0	 & 165	 & 0.5	 & 16	 & 4.7	 & 3.2	 & 400	 & 1024	 & 320	 & 4.1	 & 1.1	 & 0.50	 & 1.8	 & 16	 & 5500	 & 77	 & 8.79 $\pm$ 0.03	 & 8.5 $\pm$ 0.5 \\ 
	ngroove32	 & 17.0	 & 165	 & 0.5	 & 32	 & 4.7	 & 3.2	 & 400	 & 1024	 & 320	 & 4.1	 & 1.1	 & 0.50	 & 1.8	 & 16	 & 5500	 & 65	 & 8.27 $\pm$ 0.05	 & 2.1 $\pm$ 0.7 \\ 
	
\hline
\multicolumn{19}{c}{\textbf{Herringbone Riblets: $\boldsymbol{\alpha = 0\degree}$, parallel variant}} \\
\hline
	
	ngroove0.5	 & 17.0	 & 0	 & 0.5	 & 0.5	 & 4.6	 & 3.1	 & 400	 & 1024	 & 320	 & 4.0	 & 1.1	 & 0.50	 & 1.8	 & 16	 & 5500	 & 81	 & 7.49 $\pm$ 0.03	 & -7.6 $\pm$ 0.5 \\ 
	ngroove1	 & 17.0	 & 0	 & 0.5	 & 1	 & 4.6	 & 3.1	 & 400	 & 1024	 & 320	 & 4.0	 & 1.1	 & 0.50	 & 1.8	 & 16	 & 5500	 & 71	 & 8.06 $\pm$ 0.03	 & -0.5 $\pm$ 0.5 \\ 
	ngroove4	 & 17.0	 & 0	 & 0.5	 & 4	 & 4.6	 & 3.1	 & 400	 & 1024	 & 320	 & 4.0	 & 1.1	 & 0.50	 & 1.8	 & 16	 & 5500	 & 75	 & 7.77 $\pm$ 0.04	 & -4.1 $\pm$ 0.6 \\ 
	ngroove16	 & 17.0	 & 0	 & 0.5	 & 16	 & 4.6	 & 3.1	 & 400	 & 1024	 & 320	 & 4.0	 & 1.1	 & 0.50	 & 1.8	 & 16	 & 5500	 & 78	 & 7.64 $\pm$ 0.08	 & -5.7 $\pm$ 1.0 \\ 

\hline

    \end{tabular}
    }
  \label{tab:simulation_parameters}%
  \end{minipage}
\end{table*}%

\end{document}